\documentclass[aps,prd,nofootinbib,superscriptaddress,preprintnumbers,balancelastpage,longbibliography]{revtex4-1}

\usepackage{appendix}
\usepackage[utf8]{inputenc}
\usepackage{amsmath,amssymb,mathtools,bm}
\usepackage{graphicx, color, hepunits}
\usepackage[dvipsnames]{xcolor}
\usepackage{float}
\usepackage{physics} 
\usepackage{multirow}
\usepackage{cancel}
\usepackage{tikz-feynman}
\usepackage{hyperref}

\hypersetup{
	colorlinks=true,       
	linkcolor=blue,        
	citecolor=blue,        
	filecolor=magenta,     
	urlcolor=blue          
}
\usepackage[utf8]{inputenc}
\usepackage[english]{babel}
\usepackage{soul}

\usepackage{tensor}
\usepackage{slashed}
\usepackage[normalem]{ulem}

\begin{document}
	
	\title{Low-Frequency Gravitational Bremsstrahlung and Memory in a Medium}
	
	\author{Diego Blas}
	\affiliation{Institut de Fisica d’Altes Energies (IFAE), The Barcelona Institute of Science and Technology, Campus UAB, 08193 Bellaterra (Barcelona), Spain}
	\affiliation{Instituci\'{o} Catalana de Recerca i Estudis Avan\c{c}ats (ICREA), Passeig Llu\'{i}s Companys 23, Barcelona, 08010, Spain}
	\author{Silvia Gasparotto}
	\affiliation{CERN, Theoretical Physics Department, Esplanade des Particules 1, Geneva 1211, Switzerland}
	\author{Hitoshi Murayama}\thanks{Hamamatsu Professor}
	\affiliation{Department of Physics, University of California, Berkeley, CA 94720, USA}
	\affiliation{Leinweber Institute for Theoretical Physics, University of California, Berkeley, CA 94720, USA}
	\affiliation{Kavli Institute for the Physics and Mathematics of the Universe (WPI), University of Tokyo, Kashiwa 277-8583, Japan}
	\affiliation{Ernest Orlando Lawrence Berkeley National Laboratory, Berkeley, CA 94720, USA}
	\author{Jan Schütte-Engel}
	\affiliation{Department of Physics, University of California, Berkeley, CA 94720, USA}
	\affiliation{RIKEN iTHEMS, Wako, Saitama 351-0198, Japan}
	\author{Jann Zosso}
	\affiliation{Center of Gravity, Niels Bohr Institute, Blegdamsvej 17, 2100 Copenhagen, Denmark}

	\date{\today}
	
	\begin{abstract}

		Gravitational radiation from the early Universe offers a powerful window into physics beyond the Standard Model (SM). In this paper, we investigate the low-frequency tail of gravitational radiation produced by the decay of massive particles. The energy spectrum in Minkowski space-time, $dE/d\omega$, exhibits two distinct regimes: a flat bremsstrahlung/memory spectrum above a characteristic cutoff frequency and a suppressed quadratic frequency scaling below it. The mean free path of the decay products determines the cutoff frequency. 
		We calculate $dE/d\omega$ in two independent but equivalent ways: using Weinberg's soft graviton theorem and by solving the equations of motion. This further establishes the connection between Weinberg's soft graviton theorem and the gravitational memory formula, which we generalize to incorporate the effects of final-state scattering. 
		We then compute the cosmological stochastic gravitational energy spectrum $h^2\Omega_{\rm h}$ generated by many particle decays in the early Universe. 
		We demonstrate that $h^2\Omega_{\rm h}$ exhibits linear scaling with frequency above the cutoff frequency, transitioning to cubic scaling below it. Our findings reveal a suppression of the low-frequency signal compared to a naive linear extrapolation, with implications for detection prospects.
		Our results also reveal the connection between bremsstrahlung and memory in the context of cosmological stochastic gravitational energy spectra from the early Universe.
		
	\end{abstract}
	\maketitle

	\section{Introduction}%
	
	The first detection of gravitational waves (GWs)~\cite{LIGOScientific:2016aoc} has ushered in a new era in our ability to observe phenomena in our Universe and probe new physics. Today, black hole and neutron star mergers are routinely detected with laser interferometers on Earth~\cite{LIGOScientific:2025slb,LIGOScientific:2025hdt, LIGOScientific:2025yae} and the detection rate is expected to grow substantially as the sensitivity of current detectors improves and next-generation instruments~\cite{Hild:2010id,Evans:2023euw,Karnesis:2022vdp} come online.

	We can currently detect GWs in the Hz-kHz frequency range with laser interferometers on Earth.  Upcoming space-based interferometers, such as LISA~\cite{Karnesis:2022vdp}, will extend this range down to frequencies of order $10^{-4}\,\mathrm{Hz}$ \cite{LISA:2024hlh}. Pulsar timing arrays (PTAs) can probe GWs in the nHz regime~\cite{Sazhin:1978myk,Detweiler:1979wn,Hellings:1983fr,NANOGrav:2023gor}. Furthermore, atom interferometers~\cite{Dimopoulos:2007cj,Coleman:2018ozp,Badurina:2019hst,AEDGE:2019nxb} and satellite trackers~\cite{Blas:2021mqw,Foster:2025csl,Blas:2026xol} have the potential to fill the sensitivity gaps between LIGO and LISA, and between LISA and PTAs, respectively. 
	
	We also expect GW signals with frequencies above the kHz regime~\cite{Aggarwal:2025noe}. 
	Since the early Universe may have reached temperatures far exceeding the energy scales accessible to terrestrial experiments, unknown particles or processes at these scales may be detectable if they can generate distinctive GW signatures that, owing to the weakness of gravity, may still be around today. 
	Examples of signals peaking at frequencies above kHz include those that may be generated by first-order phase transitions~\cite{Witten:1984rs,Hogan:1986qda}, cosmic defects~\cite{Damour:2000wa,Damour:2001bk}, inflation~\cite{Grishchuk:1975abc,Starobinskii:1979abc,RUBAKOV1982189,1983PhLB..125..445F}, inflaton annihilation~\cite{Ema:2015dka,Ema:2016hlw,Ema:2020ggo}, preheating~\cite{Khlebnikov:1997di,Lozanov}, or thermal fluctuations in the primordial plasma~\cite{Weinberg:1972kfs,Ghiglieri:2015nfa,Ghiglieri:2020mhm,Ringwald:2020ist,Ghiglieri:2022rfp,Ghiglieri:2024ghm,Drewes:2023oxg}. See Ref.~\cite{Aggarwal:2025noe} for a review of additional sources.

	In this paper, we focus on the gravitational signal from the decay of massive particles in the early universe. One example is the decay of the inflaton into the particles that eventually make up our Universe; see Refs.~\cite{Nakayama:2018ptw,Huang:2019lgd,Ghoshal:2022kqp,Barman:2023ymn,Barman:2023rpg,Bernal:2023wus,Tokareva:2023mrt,Xu:2024fjl,Hu:2024awd,Hu:2024bha,Jiang:2024akb,Inui:2024wgj,Xu:2025wjq,Cline:2026jra} for previous calculations of the resulting cosmological stochastic gravitational energy spectrum. Further examples include the decay of dark matter particles~\cite{Allen:2019hnd} or the decay of right-handed neutrinos during leptogenesis~\cite{Murayama:2025thw}, see also Refs.~\cite{Datta:2024tne}.
	Most previous studies have focused on the high-frequency ($\gg\,$kHz) peak of the gravitational signal and extrapolated the low-frequency part of the spectrum with a linear dependence on frequency.\footnote{Some works have noted that a low-frequency cutoff should exist below which the linear scaling breaks down; see, for example, Refs.~\cite{Ghoshal:2022kqp,Tokareva:2023mrt}.} 
	While there is a worldwide effort to develop detectors sensitive to frequencies above the kHz regime, see for example Refs.~\cite{Aggarwal:2020umq,Berlin:2021txa,Domcke:2022rgu,Bringmann:2023gba,Kahn:2023mrj,Domcke:2024mfu,Carney:2024zzk,Domcke:2024eti,Capdevilla:2024cby}, and Ref.~\cite{Aggarwal:2025noe} for a review, one motivation for this work is to thoroughly investigate the low-frequency tail of the cosmological stochastic gravitational energy spectrum and to assess whether it may be easier to detect the low-frequency tail with detectors operating below the kHz regime.

	In Sec.~\ref{sec:Energy_spectrum}, we calculate the low-frequency tail of the gravitational energy spectrum, $dE/d\omega$, produced by a massive particle decaying into two massless particles in flat spacetime. Our calculation is based on the classical equations of motion, and in Sec.~\ref{sec:bremsstrahlung_and_memory} we show that it agrees with a calculation based on Weinberg's soft graviton theorem.  While calculating the energy spectrum, we incorporate the effects arising when the decay products do not travel to infinity: we assume that they are scattered by surrounding plasma particles. This effect gives rise to a natural cutoff frequency, below which $dE/d\omega$ transitions from a constant to a quadratic scaling with frequency. 
	In order to arrive at this result, we incorporate medium effects into both calculations, i.e., the one based on Weinberg's soft graviton theorem~\cite{Weinberg:1965nx} and the one that solves the classical equations of motion. Based on the latter, we show that the gravitational signal generates memory and that the well-known memory formula~\cite{Zeldovich:1974gvh,Braginsky:1985vlg,Braginsky:1987kwo} is modified by medium effects below the cutoff frequency.
	
	Finally, in Sec.~\ref{sec:Cosmology}, we compute the stochastic gravitational energy spectrum, $h^2\Omega_{\rm h}$, generated by many particle decays in an expanding Universe, again with particular emphasis on the low-frequency tail. We show that below the cutoff frequency, the spectrum transitions from a linear to a cubic scaling with frequency. The spectrum can be computed in two independent but equivalent ways: by solving the Boltzmann equations using Weinberg's soft graviton theorem, or by solving the equations of motion for the metric perturbation and subsequently summing over many particle decays and the gravitational energy they emit in the early Universe. Our calculation clarifies the connection between cosmological stochastic gravitational energy spectra computed using standard particle-physics methods and gravitational memory. We conclude in Sec.~\ref{sec:conclusion}.

	Throughout the paper we use the metric convention $\eta_{\mu\nu}={\rm diag}(1,-1,-1,-1)$ and natural units $\hbar=c=1$. Four-vectors are written with non-bold symbols, i.e., $p^\mu$, $x^\mu$.
	Momentum three-vectors are denoted by bold symbols, i.e., $\bm{p}$, with magnitude $|\bm{p}|$, while spatial three vectors are written as $\bm{x}$, with magnitude $r= |\bm{x}|$. Three- and four-vector inner products are indicated with a dot. 
	We adopt the Fourier-transform convention $f(\bm{k},\omega)= \int d^4x \, e^{i\omega t} e^{-i\bm{k}\cdot\bm{x}}\, f(\bm{x},t)$.

	\section{Gravitational Energy Spectrum}\label{sec:Energy_spectrum}%

\begin{figure}
		\centering
		\includegraphics[width=0.36\textwidth]{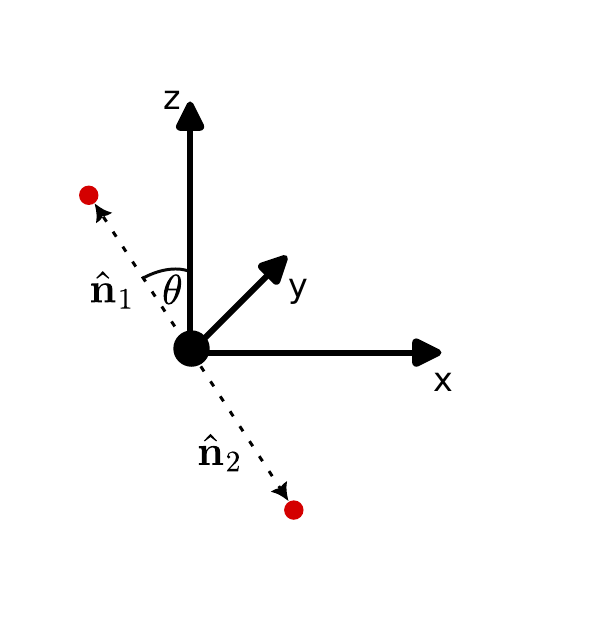}
		\includegraphics[width=0.63\textwidth]{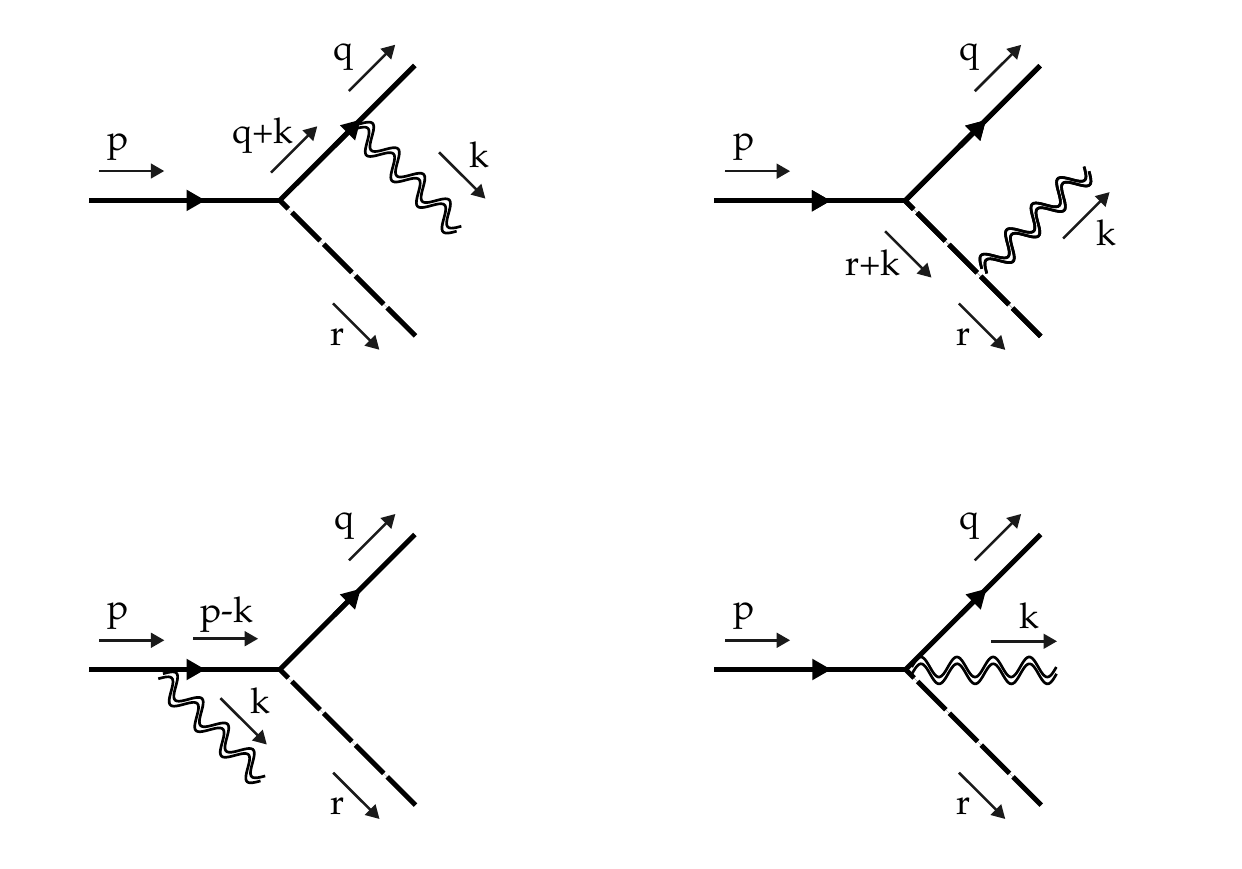}
		\caption{Left: Decay of a massive particle (black) into two massless particles (red). Right: Feynman diagrams for the decay of a lepton into a lepton and Higgs particle. Four diagrams emit a graviton as soft bremsstrahlung.}
		\label{fig:setup}
	\end{figure}

	In this section, we consider the decay of a particle of mass $M$ into two massless daughter particles, labeled by $a=1,2$, see Fig.~\ref{fig:setup} (Left). Our goal is to derive the gravitational energy emitted per frequency interval.
	We work in the rest frame of the decaying particle, using coordinates $(t,x,y,z)$, and take the decay to occur at $t=0$. The corresponding energy-momentum tensor is approximated as 
	\begin{eqnarray}
		T^{\mu\nu}(\bm{x},t)= \frac{p^\mu p^\nu}{p^0} \delta(\bm{x})\,\Theta(-t)\, e^{t/\tau}+\sum_a \frac{p_a^\mu p_a^\nu}{p_a^0}\,\delta(\bm{x}-\hat{\bm{n}}_at)\,\Theta(t) e^{-t/t_\star}.
		\label{eq:Tmunu}
	\end{eqnarray}
	The first term in Eq.~\eqref{eq:Tmunu} describes the decaying massive particle, with four-momentum $p^\mu=(M,0,0,0)$ and lifetime $\tau$.
	The second term describes the two massless decay products\footnote{ In Eq.~\eqref{eq:Tmunu} we are explicitly considering decay products propagating at the speed of light. If their velocity is decreased, this also has an impact on the final result that we do not consider in this work.}. They have four-momenta $p^\mu_a=M/2(1,\hat{\bm{n}}_a)$, where $\hat{\bm{n}}_1=-\hat{\bm{n}}_2\equiv \hat{\bm{n}}$ and are produced instantaneously. The time scale $t_\star$ characterizes how the decay products interact with a surrounding plasma with temperature $T<M$. 
	The time scale $t_\star$ is the mean free time of the daughter particles to scatter off a plasma particle and undergo an order-one change in momentum.
	We also define the rate $\omega_\star\equiv 1/t_\star$, for which we derive explicit expressions in Appendix~\ref{app:cutoff_frequency} for a concrete interaction, showing its dependence on the composition of the plasma.
	Note that the tensor in Eq.~\eqref{eq:Tmunu} is by itself not required to satisfy energy-momentum conservation. This is because it describes only the decay products which exchange energy and momentum with the surrounding medium.

	The Fourier-Transformation of the energy-momentum tensor is:
	\begin{eqnarray}
		T^{\mu\nu}(\bm{k},\omega)=
		\frac{p^\mu p^\nu}{p^0} \,  \frac{1}{i\omega+1/\tau} 
		-
		\sum_a \frac{p_a^\mu p_a^\nu}{p_a^0}\, \frac{1}{i\omega -i\bm{k}\cdot\hat{\bm{n}}_a-1/t_\star}.
		\label{eq:Tmunu_Fourier}
	\end{eqnarray}
	Notice that the time scales $\tau$ and $t_\star$ in Eq.~\eqref{eq:Tmunu} now yield the correct causal prescription when we calculate the metric perturbation in real space,  as we demonstrate in Appendix~\ref{app:metric_diploe_approximation}.
	Here, we provide them with a physical interpretation and incorporate them directly into the energy-momentum tensor from the beginning.

	In the following we expand the metric perturbation around flat space-time $g_{\mu\nu}=\eta_{\mu\nu}+h_{\mu\nu}$, $h_{\mu\nu}\ll1$ which, in De Donder (DD) gauge ($\partial^\nu\bar{h}_{\mu\nu}=0$),  gives the equations of motion
	\begin{eqnarray}
		\Box \bar{h}_{\mu\nu}(\bm{x},t)=16\pi G \,T_{\mu\nu}(\bm{x},t),
		\label{eq:Wave_eq_hmunu}
	\end{eqnarray}
	where $G=1/m_{\rm pl}^2$ is Newton's constant, $\bar{h}_{\mu\nu}=h_{\mu\nu}-\frac{1}{2}h\eta_{\mu\nu}$, with $h=\eta^{\alpha\beta} h_{\alpha\beta}$ and the inverse relation is $h_{\mu\nu}=\bar{h}_{\mu\nu}-\frac{1}{2}\bar{h}\eta_{\mu\nu}$. As mentioned, we neglect the energy-momentum tensor of the plasma, required by  $\partial^\nu\bar{h}_{\mu\nu}=0$, because here we concentrate entirely on gravitational signatures produced from the particle decay.\footnote{Soft-gravitons that are emitted due to thermalization processes or thermodynamic properties of the plasma do not depend on the decay process and should be treated with the methods of Refs.~\cite{Drewes:2023oxg,Ghiglieri:2020mhm,Ghiglieri:2015nfa}.}
	The solution of Eq.~\eqref{eq:Wave_eq_hmunu}, for large distance from the source reads \cite{Weinberg_Gravitation,Maggiore}
	
	\begin{eqnarray}
		\bar{h}_{\mu\nu}(\bm{x},t)=\frac{4G}{r}\,  \int\frac{d\omega}{2\pi} e^{-i\omega (t-r)} T_{\mu\nu}(\bm{k}=\omega\hat{\bm{x}},\omega),
		\label{eq:hmunu_far_field_final}
	\end{eqnarray}
	where, in particular, we have made the dipole approximation: $|\bm{x}-\bm{x}'|\simeq r -\hat{\bm{x}}\cdot\bm{x}'$, with $r=|\bm{x}|$. The  emitted gravitational energy is
	\begin{eqnarray}
		\frac{dE}{d\omega}= 
		\frac{G}{2\pi^2 }\, \, \int_{S} d\hat{\Omega} \,   \,  
		\omega^2     
		T^{\rm TT}_{ij}(\bm{k}=\omega\hat{\bm{x}},\omega) 
		T^{{\rm TT}\, ij *}(\bm{k}=\omega\hat{\bm{x}},\omega) ,
		\label{eq:dEdomega_main_text}
	\end{eqnarray}
	where $S$ is a unit sphere around the origin, the angular integral is over the variables $\hat{\bm{x}}=(\sin\hat{\theta}\cos\hat{\phi},\sin\hat{\theta}\sin\hat{\phi},\cos\hat{\theta})$ and the TT part of a tensor is extracted as $T^{\rm TT}_{ij}(\bm{k},\omega)=\Lambda_{ijab}(\hat{\bm{k}})T_{ab}(\bm{k},\omega)$, see Appendix~\ref{app:Tmunu_gravity} for the definition of $\Lambda$. This is a well-known result, e.g., Ref.~\cite{Weinberg_Gravitation}, that we re-derive carefully in Appendix.~\ref{app:Tmunu_gravity} to outline all underlying assumptions.

	Plugging in the expression for the energy-momentum tensor from Eq.~\eqref{eq:Tmunu_Fourier} into Eq.~\eqref{eq:dEdomega_main_text} yields:
	\begin{eqnarray}
		\frac{dE}{d\omega}&=& 
		\left(\frac{M^2}{m_{\rm pl}^2}\right)\, 
		f(\omega,\omega_\star),
		\label{eq:dEdomega_general_evaluated_main_text}
	\end{eqnarray}
	with
	\begin{eqnarray}
		f(\omega,\omega_\star)\equiv\frac{1}{8\pi }\left[-2\frac{\omega_\star}{\omega} \left(3\left(\frac{\omega_\star}{ \omega}\right)^2 +4\right) \tan ^{-1}\left(2 \frac{\omega}{\omega_\star} \right)+8  \left(\left(\frac{\omega_\star}{\omega}\right)^2+1\right)+\left(\frac{\omega_\star}{\omega}\right)^4 \log \left(4 \left(\frac{\omega}{\omega_\star}\right)^2+1\right)
		\right].
	\end{eqnarray}
	Note that in deriving the expression for $dE/d\omega$, the contribution from the first term in the energy-momentum tensor in Eq.~\eqref{eq:Tmunu_Fourier}, which depends on $\tau$ and originates from the decaying particle, vanishes since the spatial components of its momentum are zero.  Physically, this is expected: a particle at rest does not contribute to the emitted gravitational energy. For the second term in the energy-momentum tensor, corresponding to the massless decay products, we use $p_{ai}p_{aj}/(p_a^0)=(M/2) \hat{n}_{ai} \hat{n}_{aj}$. 
	We extracted the TT part and used the identity $\Lambda_{ijnm}(\hat{\bm{x}})\Lambda_{ijlk}(\hat{\bm{x}})= \Lambda_{nmij}(\hat{\bm{x}})\Lambda_{ijlk}(\hat{\bm{x}})=\Lambda_{nmlk}(\hat{\bm{x}})$. Without loss of generality we chose $\hat{\bm{n}}=\hat{\bm{e}}_z$, which gives $\hat{\bm{x}}\cdot \hat{\bm{n}}=\cos\hat{\theta}$ and we furthermore used that $\Lambda_{3333}(\hat{\bm{x}})=1/2 \,(1-\cos^2\hat{\theta})^2$.
	
	The function $f(\omega,\omega_\star)$ has a simple expression in the two limits of interest:
	\begin{eqnarray}
		f(\omega,\omega_\star) =
		\begin{cases}
			\frac{1}{\pi} & \omega \gg \omega_\star\\
			\frac{8}{15 \pi} (\omega/\omega_\star)^2 & \omega \ll \omega_\star
		\end{cases}.
		\label{eq:dEdomega_limits}
	\end{eqnarray}
	For frequencies $\omega>\omega_{\star}$ the energy spectrum becomes constant,\footnote{Note also that the constant regime does not extend to arbitrarily large frequencies and breaks down around $\omega \simeq M/2$. In this work, we focus exclusively on the low-frequency regime.} while it gets suppressed quadratically with frequency at lower frequencies. It is precisely this behavior that leads to an additional suppression of the cosmological stochastic gravitational energy spectrum from particle decays in the early universe. While the constant scaling at small frequencies is well known,  see e.g. Refs.~\cite{Weinberg_Gravitation,Smarr:1977fy,Turner:1978jj}, the extra suppression below a cutoff frequency arising from the interaction of the decay products with the medium has not, to our knowledge, been discussed in the context of signals from the early Universe. 
	The scattering process of the decay products in a medium can further suppress the emission and propagation of soft gravitational radiation \cite{Baier:1997wa,Klein:1998du}. This suppression is not relevant for the process we study since we deal with the gravitational radiation from the decay processes only.

	\section{Soft Graviton Theorem and Gravitational Memory}\label{sec:bremsstrahlung_and_memory}%
	
	In the previous section, we have calculated the soft gravitational energy emission spectrum from a particle decay process, see also Refs.~\cite{Turner:1978jj,Weinberg_Gravitation}, which however did not consider medium effects. We used the classical equations of motion to compute the metric perturbations, which we then used to compute the energy spectrum of the emitted gravitational radiation. In this section we connect the classical calculation with Weinberg's soft graviton theorem and with gravitational memory.

	\subsection{Soft graviton theorem}\label{sec:softgr}
	
	Weinberg has shown in his seminal paper~\cite{Weinberg:1965nx} that in the limit of soft graviton emission, the total amplitude of the process including gravitons factorizes and is just the amplitude without soft gravitons times a simple factor. To account for in medium effects,   we phenomenologically extend Weinberg's soft graviton theorem \footnote{
		Note also that the extension of Weinberg's result in the presence of a gravitational wave background was recently achieved in Ref.~\cite{Ai:2025xla}.} 
	\begin{eqnarray}
		A_\lambda=A_0 \left(\frac{\kappa}{2}\right)\,\sum_{a} \eta_a \frac{p_a^\mu p_a^\nu}{p_a\cdot k+\Sigma_a} \epsilon_{\mu\nu}^{\lambda *}(\bm{k}),
		\label{eq:soft_graviton_theorem}
	\end{eqnarray}
	where $\kappa=\sqrt{32\pi G}$, $A_0$ is the amplitude without graviton emission (including soft virtual gravitons relevant for the infrared behavior in vacuum), the sum runs over all external states, $\eta_a=+1$ for outgoing particles and $\eta_a=-1$ for incoming particles, $\epsilon_{\mu\nu}^{\lambda *}(\bm{k})$ is the graviton polarization tensor, where $\lambda$ is the graviton's polarization and $\bm{k}$ the graviton momentum. Finally, 
	for the additional part in the denominator in Eq.~\eqref{eq:soft_graviton_theorem}, the effective medium term that enters into dressed propagators has to be complemented by the dressing of the interaction, see also Refs.~\cite{Wang:2002nba,Arnold:2002zm,Arnold:2002ja,Gagnon:2006hi, Jeon:1994if,Thoma:2008my}. Still, one can phenomenologically understand how to interpret $\Sigma_a$ by considering a toy model of Poisson processes with rate $\Gamma$ that affect a quantity of interest (let's call it $Q$), changing it as $Q\to \alpha Q$ after the process. The average time evolution of  $Q$ is then
	\begin{equation}
		\langle Q(t)\rangle=e^{-\Gamma t}\sum_{n=0}^\infty    \frac{(\Gamma t)^n}{n!}{\alpha}^n  Q_0 =e^{-\Gamma(1-\alpha)t}Q_0,
	\end{equation}
	where $n$ stands for the number of processes (interactions), with $P_n=(\Gamma t)^n/(n!)$ being the probability of having $n$ of them. The average is obtained by summing over all possible outcomes. The Fourier transform of the previous expression then generates a structure that is similar to the one that we introduced in Eq.~\eqref{eq:soft_graviton_theorem}.
	Hence, when considering the way in which the soft terms of \cite{Weinberg:1965nx} (Eq.~\eqref{eq:soft_graviton_theorem} at $\Sigma_a=0$) are modified by the medium, we expect
	\begin{equation}
		\Sigma_a\approx i E_a\Gamma_{\rm rel}\approx i (M/2)\Gamma_{\rm rel},
	\end{equation}
	where $\Gamma_{\rm rel}$ is the relaxation rate of the energy-momentum in the thermal plasma and $E_a$ is the energy. A key aspect is that it does not correspond to the scattering rate, but to the rate of momentum relaxation. In the scenario when a massive particle decays into two massless particles $E_a\approx M/2$. 
	As expected, adding the medium provides a natural regularization of the infrared behaviour of the amplitude, independently of the cancellation from the virtual gravitons from the $A_0$ part \cite{Weinberg:1965nx}.

	In the following, we focus on the scenario in which a fermion decays into a massless scalar and a fermion, cf. Fig.~\ref{fig:setup} (Right).  Note, however, that in the soft-graviton limit the result is independent of whether the final-state particles are fermions or bosons. We will ignore the corrections from virtual soft-gravitons, as they are only relevant in the ultra-soft limit, already cut off by $\Sigma_n$ (and also because we do not need them to compute the quantity of interest, see below).
	We use the Feynman rules derived in Refs.~\cite{Choi:1994ax,Murayama:2025thw}. The amplitude for the diagram in the bottom right in Fig.~\ref{fig:setup} (Right) vanishes because the vertex is proportional to $\eta_{\mu\nu}$, cf. Ref.~\cite{Murayama:2025thw} and contracting with the polarization tensor gives zero. Furthermore, the amplitude for the diagram in the bottom left in Fig.~\ref{fig:setup} (Right) vanishes when we contract it with the polarization tensor. This is in line with what we have seen in the classical calculation. The contribution to the radiated gravitational energy from the decaying particle, which is at rest, vanishes. 
	The squared matrix element summed over the physical graviton polarizations is
	\begin{eqnarray}
		|\mathcal{M}|^2=\sum_{\lambda}|A_\lambda|^2,
		\label{eq:M_sq}
	\end{eqnarray}
	where we evaluate the matrix element with momenta in the rest frame of the decaying particle, i.e.,
	$p^\mu=M(1,0,0,0),$ $k^\mu=\omega(1,0,0,1),$ $q^\mu=\omega_L(1,\sin\theta,0,\cos\theta )$ and $r^\mu= p^\mu-q^\mu- k^\mu$.
	The choices of $\bm{k}$ along the $z$-axis and $\bm{q}$ in the $xz$-plane are made without loss of generality. The angle $\theta$ is fixed $\cos\theta = (M^2-2M(\omega+\omega_L)+2\omega \, \omega_L)/(2\omega\omega_L)$ and for simplicity we assume $\Sigma_q=\Sigma=\Sigma_r$.
	The spin-averaged squared amplitude for the decay of a massive lepton into a massless lepton and a Higgs is $|A_0|^2=|Y|^2\,M \omega_L$
	where $Y$ is the lepton-lepton-Higgs Yukawa coupling~\cite{Murayama:2025thw}. As mentioned, we ignored the contribution from soft virtual gravitons and note that for sufficiently small graviton momentum, the $\Sigma$ term in the denominator of Eq.~\eqref{eq:M_sq}  regularizes the energy spectrum.
	
	Given the decay rate $\Gamma_0$ of the process without graviton emission, the quantity $\frac{1}{\Gamma_0}\frac{d\Gamma}{d \omega}$ represents the probability, conditioned on the parent particle decaying, that a graviton is emitted in the interval $\omega,\omega+d\omega$. Multiplication by the graviton energy, therefore, gives the average gravitational energy emitted per frequency interval
	\begin{eqnarray}
		\frac{dE}{d\omega}\equiv \omega\frac{d}{d\omega}\,\left(\frac{\Gamma}{\Gamma_0}\right).
	\end{eqnarray}

	When the $\Sigma$ term in the denominator inside Eq.~\eqref{eq:M_sq} is subdominant, we recover the first case of Eq.~\eqref{eq:dEdomega_limits}. On the other hand, if the graviton momentum is sufficiently small such that the $\Sigma$ term dominates, we obtain
	\begin{eqnarray}
		\frac{dE}{d\omega} 
		\approx\frac{2}{15\pi}\,  \frac{M^2}{m_{\rm pl}^2}  \, \frac{M^2 }{ |\Sigma|^2 } \omega^2.
	\end{eqnarray}
	Comparing the result with the second case in Eq.~\eqref{eq:dEdomega_limits}, one finds
	\begin{eqnarray}
		\omega_\star=2|\Sigma|/M\approx \Gamma_{\rm rel},
	\end{eqnarray}
	as expected from the interpretation of $\Gamma_{\rm rel}$.
	In summary, we have shown that the graviton emission spectrum can be calculated in two independent ways that lead to the same result. First, one can solve the classical equations of motion and compute the gravitational energy spectrum from the energy-momentum tensor describing the decay process (Sec.~\ref{sec:Energy_spectrum}). In this section, we have shown that the same soft-graviton energy spectrum can also be derived using Feynman diagrams together with Weinberg's soft graviton theorem.

	\subsection{Gravitational memory}
	
	Gravitational memory refers to the permanent displacement between two test masses after gravitational radiation has passed through them. It was first discovered in the context of astrophysical energy bursts, including close encounters between astrophysical objects, supernova explosions, neutrino bursts and gamma-ray bursts., cf. Refs.~\cite{Zeldovich:1974gvh,Braginsky:1985vlg,Braginsky:1987kwo,Turner:1978jj,Epstein:1978gvh,Burrows:1995bb,1997A&A...317..140M,Sago:2004pn,Ott:2008wt,Murphy:2009}. Such phenomena were later classified as \textit{linear memory}, as it was subsequently realized that gravitational energy bursts themselves can also induce a permanent displacement, which became known as \textit{non-linear memory} (see Refs.~\cite{Christodoulou:1991cr,Ludvigsen:1989cr,Blanchet:1992br,Thorne:1987sdb,Thorne:1992sdb,Caldwell:2025tfu}).
	While the initial studies on gravitational memory focused on astrophysical scenarios, the effect can also arise from microscopic particle decays and scattering events (see, e.g., Refs.~\cite{Tolish:2014oda,Tolish:2014bka}). We will focus on such scenarios in this work.

	In the following, we first review gravitational memory in the absence of scattering effects ($t_\star\to\infty$). We begin by solving the equations of motion for the metric perturbation in the massive particle decay scenario, illustrated in Fig.~\ref{fig:setup} (Left). The full solution for the metric perturbation is
	\begin{eqnarray}
		\bar{h}_{\mu\nu}(\bm{x},t) = \frac{4G}{r} \left[\frac{p_\mu p_\nu}{p^0} \,\Theta(r-t) +\sum_{a=1,2}\frac{1}{t/r-\hat{\bm{n}}_a\cdot \hat{\bm{x}}} \frac{p_{a\mu} p_{a\nu}}{p^0_a}\Theta(t-r) \right],
		\label{eq:hmunuFull}
	\end{eqnarray}
	where we provide details of the computation in Appendix~\ref{app:details_metric}.
	
	The first term in Eq.~\eqref{eq:hmunuFull} originates from the decaying particle and is nonzero only for  $t\leq r$, which is consistent with causality since the decay happens at $r=0,\, t=0$. 
	The second term in Eq.~\eqref{eq:hmunuFull} becomes nonzero once the information about the decay causally reaches the observer. 
	If $\hat{\bm{n}}_a\cdot \hat{\bm{x}}\neq 0$, the second term is finite and peaks at $t=r$, after which the amplitude of the metric perturbation decays with time for $t>r$. The gravitational signal from the decay therefore manifests as a pulse in the metric perturbation.
	
	The metric perturbation must remain small $\bar{h}_{\mu\nu}\ll 1$ at all times. This is the case if $GM/r\ll 1$, or equivalently $\left(M/m_{\rm pl}\right)\,\left(\left( 1.61\times 10^{-35}\,{\rm m}\right)/r\right)\ll 1$, what makes it clear that for masses $M<m_{\rm pl}$ and distances $r> 1.61\times 10^{-35}\,{\rm m}$ the metric perturbation always remains perturbative.
	The second term in Eq.~\eqref{eq:hmunuFull} can in principle become non-perturbatively large when $\hat{\bm{n}}\cdot \hat{\bm{x}}\to t/r$. For $t=r$ this corresponds to approaching the particle trajectories. However, note that taking the transverse-traceless part to select out the physically gauge-invariant radiation removes this divergence.

	Using the dipole expansion expression of the metric perturbation, cf. Eq.~\eqref{eq:hmunu_far_field_final}, we recover the same result as the full solution, cf. Eq.~\eqref{eq:hmunuFull} but with $t/r=1$ in the denominator. In other words, the dipole approximation is only valid when $t$ and $r$ are sufficiently close to one another. This in fact turns out to be exactly the limit we care about: we want to extract the $1/r$ radiative part of the metric, which means that we have to consider large $r$. But in this limit we must also send $t\to\infty$ in order to extract the interesting part where the theta function switches on. This is known as the limit to null infinity, in which we send both $r$ and $t$ to infinity while keeping track of small but positive $u=t-r$.

	The gravitational energy flux produced by the decay process induces a so-called \textit{displacement gravitational memory} effect, i.e., the proper distance between two test masses differs before and after the passage of the gravitational radiation burst. We call $s(t)$ the proper distance between two space-time events $(t,\bm{x}_1)$ and $(t,\bm{x}_2)$. Then 
	\begin{eqnarray}
		\Delta s\equiv s(t>r)-s(t<r)=-\frac{1}{2}\frac{1}{|\Delta \bm{x}|}\Delta h_{ij}^{\rm TT}\Delta x^i \Delta x^j,
	\end{eqnarray}
	where $\Delta \bm{x}=\bm{x}_2-\bm{x}_1$. Here, $t>r$ ($t<r$) should be understood as the limit to null infinity, i.e., both $t$ and $r$ are taken to be large while $t$ remains slightly larger (smaller) than $r$. This is justified because, for large $r$, the observational window around $t=r$ becomes narrow, so that for all practical purposes $t/r\simeq 1$. In deriving the above equation, we have assumed that $\bm{x}_1$ and $\bm{x}_2$ are sufficiently close together that the metric perturbation can be treated as constant when evaluating the proper-distance integral.

	Using the dipole approximation, we find, see also Ref.~\cite{Braginsky:1987kwo}:
	\begin{eqnarray}
		\Delta h_{ij}^{\rm TT} =
		\frac{4G}{r}\,  \left[
		\sum_a \frac{p_{ai} p_{aj}}{p_a^0}\,\frac{1}{1 -\hat{\bm{x}}\cdot\hat{\bm{n}}_a}  
		-
		\frac{p_i p_j}{p^0} 
		\right]^{\rm TT}
		=
		\frac{1}{r}\, \sqrt{\frac{G}{2\pi}} \left[
		\sum_a \frac{p_{ai} p_{aj}}{p_a\cdot \tilde{k}}  
		-
		\frac{p_i p_j}{p\cdot \tilde{k}} 
		\right]^{\rm TT},
		\label{eq:DeltahijTT}
	\end{eqnarray}
	where we have defined the four-vector $\tilde{k}^\mu=1/\sqrt{32\pi G}\,(1,\hat{\bm{x}})$.
	It is directly evident that the memory formula Eq.~\eqref{eq:DeltahijTT} and Weinberg's soft graviton theorem~\cite{Strominger:2014pwa} appear to have a very similar structure. In Ref.~\cite{Strominger:2014pwa}, Strominger and Zhiboedov have derived one expression from the other (in the case $\Sigma=0$).
	
	The fact that we have solved the equations of motion to derive the memory formula makes the connection with Sec.~\ref{sec:Energy_spectrum} clear. There we have calculated the energy spectrum on the basis of the classical equations of motion. And in this section we have shown that starting from the equations of motion we can derive the memory effect. Heuristically, we can therefore conclude that the gravitational energy spectrum derived in Sec.~\ref{sec:Energy_spectrum} includes the memory effect. Note that in order to derive the memory effect formula we have used the dipole approximation, which in the limit of interest $t/r\approx 1$  captures the memory effect.

	Next, we discuss the effect on gravitational memory due to the scatterings of the decay products with a surrounding plasma, i.e., we have finite $t_\star$ or, in the language of finite width, non-zero $\Sigma$.
	Starting with the dipole expansion expression for the metric perturbation, cf. Eq.~\eqref{eq:hmunu_far_field_final} we obtain:
	\begin{eqnarray}
		\bar{h}_{\mu\nu}(\bm{x},t)=
		\frac{4G}{r}\,  \left[\frac{p_\mu p_\nu}{p^0} \,  \exp\left( \frac{t-r}{\tau}\right)\Theta(r-t) 
		+
		\sum_a \frac{p_{a\mu} p_{a\nu}}{p_a^0}\,\frac{1}{1 -\hat{\bm{x}}\cdot\hat{\bm{n}}_a} \,  \exp\left(- \frac{t-r}{t_\star(1-\hat{\bm{x}}\cdot \hat{\bm{n}}_a)}\right)\Theta(t-r) \right].
		\label{eq:humnubar}
	\end{eqnarray}
	Additional details are provided in Appendix~\ref{app:metric_diploe_approximation}. We note again that this expression is only valid when $t$ is sufficiently close to $r$.
	This equation modifies the known memory formula.
	For finite $t_\star$, it leads to a finite-retarded-time waveform for the radiative TT part of the instantaneous memory rise, whose memory contribution is suppressed by the exponential factor,
	\begin{eqnarray}
		\Delta h_{ij}^{\rm TT} (u_-, u_+)=
		\frac{1}{r}\,  \sqrt{\frac{G}{2\pi}} \left[
		\sum_a \frac{p_{ai} p_{aj}}{p_a\cdot\tilde{k}}  
		\,\exp\left(- \frac{u_+}{t_\star(1-\hat{\bm{x}}\cdot \hat{\bm{n}}_a)} \right)
		-
		\frac{p_i p_j}{p\cdot \tilde{k}} \,\exp\left(-\frac{u_-}{\tau}\right) 
		\right]^{\rm TT},
		\label{eq:DeltahijTT_damping}
	\end{eqnarray}
	where $u_+ = t-r\geq0$ is a positive number and not too large compared to $r$ and we in the limit $t,r\to\infty$. Similar  $u_-=r-t\geq 0$ with $t\to-\infty$ such that $u_- >0$. Note that in the limit $u_\pm\to\infty$, $\Delta h_{ij}^{\rm TT}$ vanishes.

	\section{Stochastic Gravitational Energy Spectrum From Many Particle Decays}\label{sec:Cosmology}%

	The cosmological stochastic gravitational energy density today from many gravitational radiation emitting events in the universe is:
	\begin{eqnarray}
		h^2\Omega_{\rm h} =h^2\frac{1}{\rho_c^0} \frac{d\rho_h^{0}}{d{\rm ln}\omega_0} = \frac{1}{\rho_c^0}  \int dz\,\,\frac{\omega}{(1+z)}\frac{dE}{d\omega}\frac{dN}{dV\,dz},
		\label{eq:phinney_main_text}
	\end{eqnarray}
	where the factor $h^2$ removes the uncertainty associated with the measurement of the Hubble parameter, $\rho_c^0$ and $\rho_h^{0}$ denote the critical and gravitational energy densities today, $\omega_0$ is the frequency today and $dN/(dV\,dz)$ is the number of gravitational radiation emitting events occurring in a co-moving volume $dV$ and redshift interval $dz$.
	The derivation does not assume that the gravitational energy spectrum produced by a single decay is monochromatic. 
	However, it relies on the assumption that the gravitational radiation events happen instantaneously. 
	Eq.~\eqref{eq:phinney_main_text} was originally derived in Ref.~\cite{Phinney:2001di}. In Appendix~\ref{app:phinney}, we re-derive this result and carefully outline all underlying assumptions.

	One example where massive particles decay in the early universe is leptogenesis, where heavy right-handed neutrinos decay into a massless lepton and a Higgs boson. The cosmological evolution proceeds as follows. Once the plasma temperature drops below the mass scale $M$, the heavy particles become non-relativistic.  These massive particles can survive for some time before decaying, basically instantaneously, at some time $t_D$. We assume that the plasma temperature at the time of decay, $T_D$, is smaller than $M$. Consequently, the momenta of the emitted decay products are larger than the typical momenta of particles in the thermal plasma.
	
	For an instantaneous decay at $t_D$ the last factor in Eq.~\eqref{eq:phinney_main_text} is:
	\begin{eqnarray}
		\frac{dN}{dV\,dz} = \frac{dN}{dV_{\rm phy}\,dt} \frac{dt}{dz} \frac{1}{(1+z)^3} = n_D   \delta(z-z_D) \frac{1}{(1+z)^3},
	\end{eqnarray}
	where we have used that $dN/(dV\,dz)$ is positive, the co-moving and physical volume elements are related by $dV = (1+z)^{3} dV_{\rm phy}$, and the number density of the heavy particles, which is defined with respect to physical volume, at time $t_D$ is $n_D$. 
	Furthermore, $dN/(dV_{\rm phy}\,dt)= n_D\delta(t-t_D)$ which was re-written with $\delta(t-t_D) =\delta(z-z_D)H_D (1+z_D)$. 
	Combining everything, we obtain 
	\begin{eqnarray}
		h^2\Omega_{\rm h} =
		h^2\Omega_\gamma^0\frac{\omega_0}{ 2 T_0}\,  \, \frac{ \, T_D}{M}  \, g_*^0 \,  \frac{dE}{d\omega}\Big|_{z=z_D},
		\label{eq:Omegah_general}
	\end{eqnarray}		
	where $\rho_\gamma^0=\pi^2/30\, \times 2\, T_0^4$ is the present-day photon energy density, with $T_0=2.34\times 10^{-13}\,{\rm GeV}$, $g_{*}^0=3.913$~\cite{Saikawa:2018rcs} and $h^2\Omega_{\gamma}^0=h^2\rho_\gamma^0/\rho_c=2.473\times 10^{-5}$. We have also used the relation $1/(1+z_D)^3=(a_D/a_0)^3=g_*^0/g_{*{\rm SM}}\, (T_0/T_D)^3$, where $g_{*{\rm SM}}=106.75$ is the number of SM relativistic degrees of freedom.
	Furthermore, we assume that the decaying particles are non-relativistic at the time of decay such that $n_D=\rho_D/M$. We also assume that, at the time of decay, the energy density of the decaying particles is equal to that of the relativistic SM plasma, $\rho_D=\rho_{\rm SM}=\pi^2/30\, g_{*{\rm SM}} T_D^4$.

	Note that Eq.~\eqref{eq:phinney_main_text} is completely general and remains valid for non-standard cosmological histories. In this work, we assume a standard thermal evolution following the decay, in which the plasma consists solely of SM particles and no significant entropy production occurs. If the subsequent cosmological evolution were modified, the relation between the redshift factor $(1+z)^{-3}$ and the plasma temperature would be altered. Consequently, $h^2\Omega_{\rm h}$ would be modified as well.
	
	Using the expression for $dE/d\omega$, cf. Eq.~\eqref{eq:dEdomega_general_evaluated_main_text}, the low frequency stochastic gravitational energy spectrum is:
	\begin{eqnarray}
		h^2\Omega_{\rm h} =1.8\times 10^{-22} \left(\frac{\omega_0}{10^6\,{\rm Hz}}\right)\,\left(\frac{T_D}{10^{12}\,{\rm GeV}}\right)\,\left(\frac{M}{10^{14}\,{\rm GeV}}\right) f(\omega_0,\omega_\star^0),
		\label{eq:Omegah_today}
	\end{eqnarray}
	where the frequencies at production $\omega_D$ and $\omega_{\star}^D$ have been redshifted to today $\omega_0=\omega_D/(1+z_D)$, $\omega_\star^0=\omega_\star^D/(1+z_D)$.
	
	In Appendix~\ref{app:cutoff_frequency}, we derive an expression for the cutoff frequency, $\omega_\star$, which is the inverse of the mean free path over which a decay product loses an order one fraction of its initial momentum. In a model with an electromagnetic-like interaction with coupling constant $e$, it is given by, cf. Eq.~\eqref{eq:omega_TM_numbers}:
	\begin{eqnarray}
		\omega_{\star}^0\approx 4.0\times 10^7\,{\rm Hz}\,\left(\frac{T_D}{10^{12}\,{\rm GeV}}\right)\,\left(\frac{10^{14}\,{\rm GeV}}{M}\right)\,\left(\frac{e}{0.5}\right)^4,
		\label{eq:omega_star_0}
	\end{eqnarray}
	where in order to get the correct pre-factor we have also evaluated the logarithm that appears in $\omega_\star^0$ at $T_D=10^{12}$ GeV, $M=10^{14}$ GeV and $e=0.5$, cf. Eq.~\eqref{eq:omega_star_QED}.
	We can compare this with the Hubble rate at the decay redshifted to today cf.~\eqref{eq:omega_TM_H_numbers}:
	\begin{eqnarray}
		\frac{H_D}{1+z_D}=1.6\times 10^5\,{\rm Hz}\,\left(\frac{T_D}{10^{12}\,{\rm GeV}}\right).
		\label{eq:H_today}
	\end{eqnarray}
	Comparing the expression with Eq.~\eqref{eq:omega_star_0} makes clear that only by pushing $M$ above $2.5\times10^{16}\,{\rm GeV}$  does $\omega_{\star}^0$ become smaller than the redshifted Hubble rate. In all other cases, the scattering cutoff remains larger. To be consistent with the assumptions of this section, $\omega_\star^0$ should be larger than the cosmological expansion at $z_D$, cf. Eq.~\eqref{eq:H_today}.

	\begin{figure}
		\centering
		\includegraphics[width=0.6\textwidth]{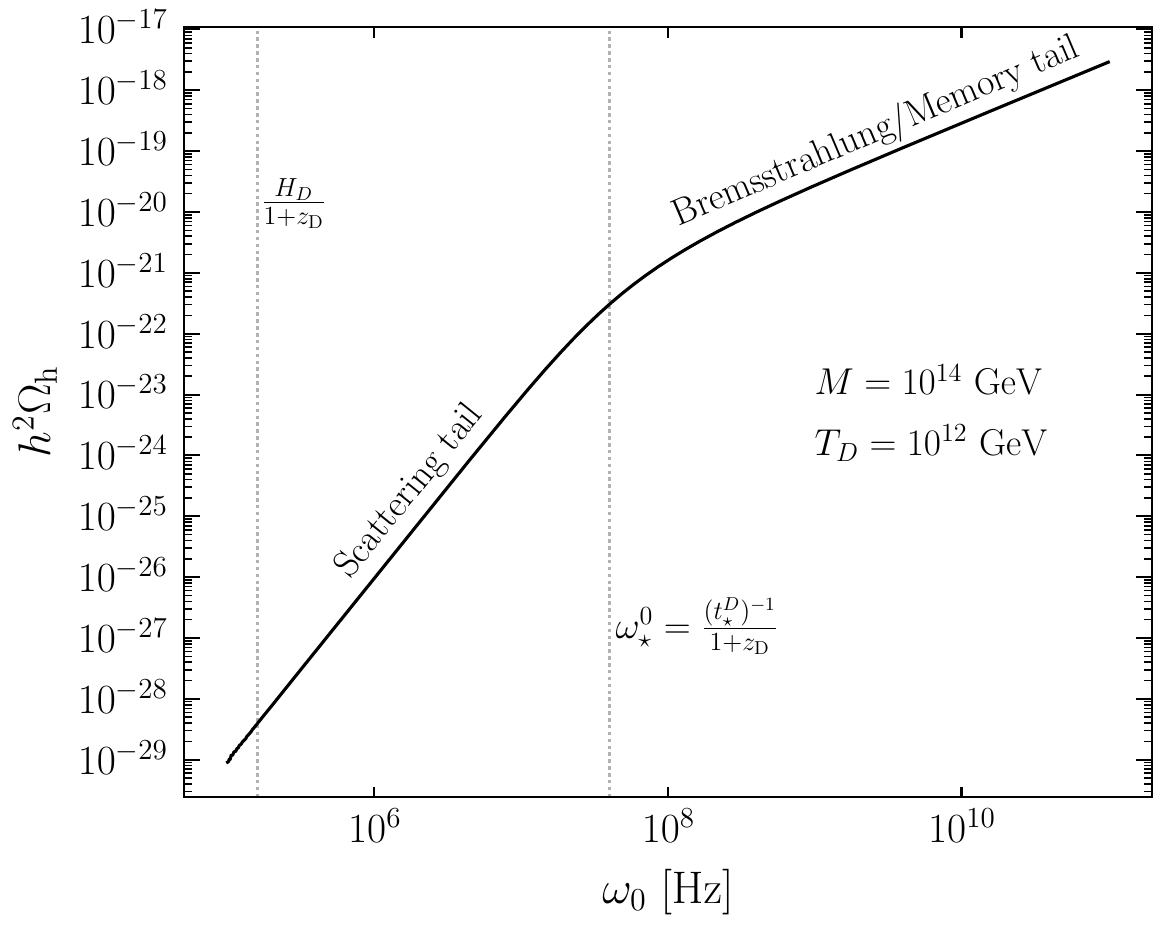}
		\caption{Cosmological stochastic gravitational energy spectrum from particles that decay at time $t_D$ in the early Universe. The spectrum is shown as a function of the frequency observed today. Above the cutoff frequency, the spectrum scales linearly with frequency (the``Bremsstrahlung/memory tail"), while below the cutoff it follows a cubic scaling. The cutoff frequency is determined by the interaction of the decay products with a surrounding plasma ($\omega_\star^0=(t_{\star}^D)^{-1}/(1+z_D)$). We also show the frequency corresponding to the Hubble rate at the time of decay, redshifted to today. The mass of the decaying particles is $M=10^{14}\,$GeV and the temperature of the surrounding thermal plasma is $T_D=10^{12}\,$GeV.  }
		\label{Fig:GW_spectrum}
	\end{figure}

	We plot the cosmological stochastic gravitational energy spectrum in Fig.~\ref{Fig:GW_spectrum}. The cutoff frequency is depicted with a dotted vertical line at $\omega_\star^0$. The other dotted vertical line represents the redshifted Hubble rate at the time of the decay. Above the cutoff frequency, the spectrum scales linearly~\footnote{Note that the linear scaling is only valid up to frequencies where $\omega$ becomes on the order of $M/2$.} with frequency (``the Bremsstrahlung/Memory tail'') whereas below it the spectrum scales as the cube of the frequency (``the scattering tail''). 
	Note that for frequencies below the Hubble scale one would in principle have to solve the equations of motion for the tensor modes in a Friedmann-Lemaitre-Robertson-Walker background. For short-lived sources, it has been shown that $h^2\Omega_{\rm h}$ also scales with frequency cubed below the Hubble scale, cf. Refs.~\cite{Caprini:2009fx,Barenboim:2016mjm,Cai:2019cdl,Ellis:2020nnr,Hook:2020phx,Guo:2020grp,Brzeminski:2022haa}. 
	For $\omega_\star^0>H_D/(1+z_D)$, we do not expect major modifications to the shown spectrum. 
	However, reducing the coupling of the daughter particles, $e$, the scattering cutoff will eventually fall below the Hubble scale. This is not relevant for SM gauge couplings but, for example, for very weakly interacting beyond the SM particles. In such a scenario, one must solve the cosmological tensor-mode equations, which we leave for future work.
	
	The cubic frequency scaling below the scattering cutoff has profound implications for the detection of the signal.\footnote{Here we refer to the signal from the decay process that generates a memory tail. As we discussed, the momentum is damped into the plasma that generates its own GW signal, though with properties that we expect to be rather independent of the decay process.} Without the cutoff, one would conclude that the gravitational energy spectrum is much larger, especially in low-frequency regimes where interferometers reach their peak sensitivity. Our results demonstrate that it will be much harder to detect these signals with either Earth-based or space-based interferometers. On the other side, this further motivates the development of detectors in the high-frequency regime where the signal is not suppressed.

\begin{figure}
		\centering
		\includegraphics[width=0.9\textwidth]{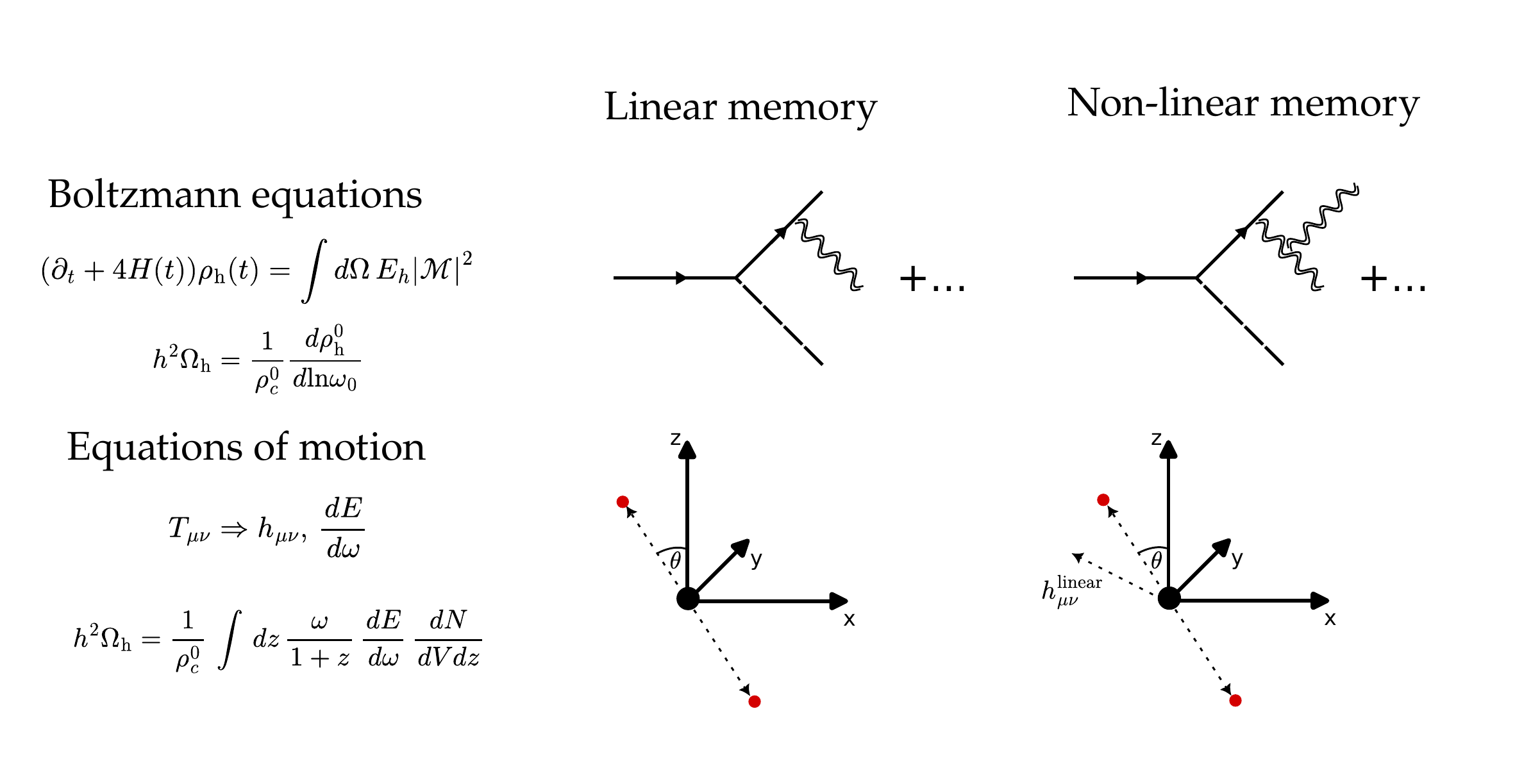}
		\caption{We illustrate two equivalent methods for computing the low-frequency tail of $h^2\Omega_{\rm h}$ from particle decays. The first relies on Feynman diagrams and Boltzmann equations; the second on the classical solution of the equations of motion and the computation of the gravitational energy spectrum $dE/d\omega$. Both methods can be used to calculate the linear and non-linear memory contributions. In the language of Feynman diagrams, linear memory corresponds to diagrams with one soft graviton in the final state, while in the language of classical equations of motion it amounts to including all particles of the decay process in the energy-momentum tensor. Non-linear memory corresponds to all Feynman diagrams that one obtains after attaching a soft graviton to all of the linear memory diagrams. In terms of classical equations of motion, computing the non-linear memory contribution requires incorporating the energy-momentum contribution from the linear memory metric perturbation $h_{\mu\nu}^{\rm linear}$ into the total energy-momentum tensor that sources the equations of motion for the non-linear memory metric perturbation. }
		\label{Fig:bremsstrahlung_memory}
	\end{figure}

	The result for $h^2\Omega_{\rm h}$ in the limit $\omega_0>\omega_{\star}^0$ agrees with the low-frequency tail derived in, for example, Ref.~\cite{Murayama:2025thw} where the production rate was computed using Feynman diagrams and embedded in a cosmological context via Boltzmann equations.\footnote{
		The result agrees with that of Ref.~\cite{Murayama:2025thw} up to a factor of two. This discrepancy arises because, in Ref.~\cite{Murayama:2025thw}, it was incorrectly assumed that the right-handed neutrino distribution function describes only a single spin state, whereas it actually accounts for both spin states. Correcting this assumption removes the discrepancy and yields exact agreement with Eq.~\eqref{eq:Omegah_general} in the limit $\omega > \omega_\star$.
	}
	We have thus demonstrated that there are two equivalent approaches for computing the low-frequency tail of $h^2\Omega_{\rm h}$. 
	The first computes $dE/d\omega$ and then applies Eq.~\eqref{eq:phinney_main_text} in order to calculate $h^2\Omega_{\rm h}$. Note that $dE/d\omega$ can be calculated directly with the metric perturbation that fulfills the classical equations of motion or equivalently with Weinberg's soft graviton theorem.
	The second relies on Boltzmann equations, where the collision term matrix element has to be calculated with Feynman diagrams. A schematic comparison of both methods is shown in the left column of  Fig.~\ref{Fig:bremsstrahlung_memory}. 
	Note that additional Feynman diagrams exist in which the soft graviton can be attached at other external lines; only one representative diagram is shown for illustration. We have established the equivalence of both methods in the case of a single massive particle decaying into two massless particles, and we expect this correspondence to hold in more general scenarios as well. 
	We find that below the cutoff frequency $\omega_\star$ the spectrum $h^2\Omega_{\rm h}$ transitions from a linear to a cubic scaling with frequency.

	We label the left column of Fig.~\ref{Fig:bremsstrahlung_memory} \textit{linear memory} because the energy-momentum tensor, cf. Eq.~\eqref{eq:Tmunu}, involves only the decaying particle and its two decay products. Furthermore, the corresponding Feynman diagrams contain exactly one graviton in the final state. In the right column, we illustrate the so-called \textit{non-linear memory} effect. In the language of Feynman diagrams, non-linear memory arises by attaching a soft graviton to each of the linear-memory diagrams, where the original graviton from the linear memory diagrams need no longer to be soft. This generates diagrams containing three-graviton vertices, which is an inherently non-linear gravitational effect; see for example Fig.~\ref{Fig:bremsstrahlung_memory} (top right).
	In the language of equations of motion, the gravitational energy spectrum from non-linear memory can be computed by adding the energy-momentum contribution of the linear memory metric perturbation $h_{\mu\nu}^{\rm linear}$ to the energy-momentum tensor and then solving the equations of motion. The resulting metric perturbation would likewise produce a permanent displacement of two test masses. 
	
	Stochastic gravitational non-linear memory spectra have previously been studied in Ref.~\cite{Unal:2025jnq}. 
	Since gravitons couple very weakly to matter, we expect the coupling $e$ to be reduced, leading to the scattering cutoff frequency being smaller than the redshifted Hubble rate. In this case, one has to solve the equations of motion in a cosmological background.
	The authors of Ref.~\cite{Unal:2025jnq} find the low frequency scaling $f^3 \log^2(1/f)$. While our work focuses primarily on linear memory, it would be interesting to investigate non-linear memory within our formalism, as interference terms could arise in Eq.~\eqref{eq:dEdomega_main_text} when the energy-momentum tensor is extended to include the linear-memory contribution.

	\section{Conclusion}\label{sec:conclusion}%
	
	In this work, we have analyzed the gravitational energy spectrum, $dE/d\omega$, produced from particle decays, taking into account the interactions of the decay products with the surrounding medium. We then applied our results to calculate the cosmological stochastic gravitational energy spectrum, $h^2\Omega_{\rm h}$, generated by  decaying particles in the early Universe.
	
	Our primary focus has been the low-frequency regime. We find that $dE/d\omega$ is frequency independent above a characteristic cutoff frequency and scales quadratically with frequency below it. This cutoff is determined by the time scale at which the decay products lose their energy-momentum as they interact with the surrounding medium. We derived this result in two independent ways. Firstly, we solved the classical equations of motion, incorporating the scattering of the decay products with a surrounding medium directly in the energy-momentum tensor. We then used the resulting metric perturbation to compute the energy spectrum $dE/d\omega$. Secondly, we derived the same result using a generalization of Weinberg's soft graviton theorem, assuming that the decay products acquire medium induced self-energies.

	Analogously, we showed that $h^2\Omega_{\rm h}$ can be calculated using two different but equivalent approaches. The first uses Boltzmann equations with a collision term derived from Weinberg's soft graviton theorem. The second sums the gravitational radiation emitted in many particle decays, characterized by the energy spectrum $dE/d\omega$.
	In a standard cosmological scenario, the low-frequency tail of $h^2\Omega_{\rm h}$ scales linearly with frequency above the cutoff frequency and cubically below it.
	This behavior has been overlooked in previous studies of gravitational radiation from particle decays in the early Universe.

	The suppression of $h^2\Omega_{\rm h}$ we describe in this work has important implications for detection prospects. Many stochastic gravitational backgrounds peak at very high frequencies, while their low-frequency tails extend into the frequency range where terrestrial and space-based interferometers are sensitive. The cubic low-frequency scaling substantially suppresses the signal in the regime where these detectors operate, and may also appear above the causal scale where a suppression of the signal is also expected.\footnote{Note that for single astrophysical sources, detectors operating below the kHz regime can in some cases be more sensitive to the low-frequency memory signal than to the high-frequency part~\cite{McNeill:2017uvq,Gasparotto:2025wok,Zosso:2026czc}. }

	We have assumed that the decay products are more energetic than the particles in the ambient plasma. Consequently, they thermalize through multiple scattering processes. These scatterings themselves generate additional gravitational radiation that was not included in our analysis. We also did not investigate interference effects, which we expect to further modify the spectrum, as occurs in the case of the Landau–Pomeranchuk–Migdal (LPM) effect~\cite{Landau:1953ivy,Migdal:1956tc,Landau:1965ksp,Baier:1996vi,Arnold:2002zm,Peigne:2008wu,Evans:2019vxr}. 
	In future work, we aim to solve the equations of motion for the tensor modes with a source term that includes interactions with the surrounding medium. This will also enable us to include damping effects of tensor modes~\cite{Weinberg:2003ur,Watanabe:2006qe,Domenech:2025bvr} as they propagate through the Universe.

	The metric perturbation generated by a particle decay process produces a permanent displacement of freely falling test masses, a phenomenon known as gravitational memory. We showed that computing $h^2\Omega_{\rm h}$ from the metric perturbation that produces gravitational memory yields the same result as the calculation based on Weinberg's soft graviton theorem and the Boltzmann equation. This correspondence clears up the connection between the terminology commonly used in particle physics and that used in the gravity community. In particular, $h^2\Omega_{\rm h}$ derived from linear memory is equivalent to one calculated with soft-graviton bremsstrahlung diagrams and Boltzmann equations.

	We also commented on non-linear memory. The corresponding contribution to $h^2\Omega_{\rm h}$ is suppressed by an additional factor of $1/m_{\rm pl}^2$. In the Feynman-diagram language, it can be calculated by attaching an additional soft graviton to all linear-memory diagrams. In the classical equations of motion picture, it arises from adding the linear-memory metric perturbation to the energy-momentum tensor used to compute the non-linear memory signal. The cutoff frequency for the non-linear memory may be lower than that of linear memory because gravitons couple only very weakly and therefore have a longer mean free path.
	
	Gravitational signals from the early Universe can provide access to particle-physics phenomena at energy scales beyond the reach of terrestrial experiments. Examples include the decay of ultra-massive right-handed neutrinos in leptogenesis and the decay of the inflaton, both of which can generate distinctive stochastic gravitational spectra. We have shown that the low-frequency tail of the spectra can be more suppressed than previously thought (note that Ref.~\cite{Drewes:2023oxg} also pointed out that $h^2\Omega_{\rm h}$ should be suppressed at low frequencies; however, the authors only discussed this suppression qualitatively). At the same time, further work is needed to incorporate the additional gravitational radiation generated during the subsequent thermalization of the decay products, cf., Refs.~\cite{Harigaya:2019tzu,Mukaida:2022bbo} in the context of dark matter production through thermalization.

	\acknowledgements
	We thank Kazunori Nakayama, Sanjay K. Reddy,  Nick Rodd, Jessie Shelton, and Bob Wald for useful discussions. We are also grateful to Marco Drewes, Yannis Georis, Juraj Klaric, and Philipp Klose for suggesting that the low-frequency spectrum can be suppressed. We thank Michael Wentzel for his collaboration during the early stages of this project. 
	JSE was supported by the National Science Foundation under cooperative agreement 202027 and by the by Japan Science and Technology Agency (JST) as part of Adopting Sustainable Partnerships for Innovative Research Ecosystem (ASPIRE), Grant Number JPMJAP2318. HM was supported by the NSF grant PHY-2515115, by the U.S. Department of Energy (DE-AC02-05CH11231), by the JSPS Grant-in-Aid for Scientific Research JP23K03382, MEXT Grant-in-Aid for Trans-
	formative Research Areas (A) 26H00401, 26A204, 26H00403, Hamamatsu Photonics, K.K, Tokyo Dome Corporation, and by the World Premier International Research Center Initiative (WPI) MEXT, Japan. 
	JZ is supported by funding from the Swiss National Science Foundation (Grant No.~222346 and 242481) and the Janggen-Pöhn Foundation. JZ additionally acknowledges support from the Simons Collaboration on Black Holes and Strong Gravity through its Long-Term Visit Program. The Center of Gravity is a Center of Excellence funded by the Danish National Research Foundation under Grant No. 184.
	This publication is part of the R\&D\&i project PID2023-146686NB-C31 funded by MICIU/AEI/10.13039/501100011033/ and by ERDF/EU.
	IFAE is partially funded by the CERCA program of the Generalitat de Catalunya.
	This work is supported by ERC grant ERC-2024-SYG 101167211. Funded by the European Union. Views and opinions expressed are, however, those of the author(s) only and do not necessarily reflect those of the European Union or the European Research Council Executive Agency. Neither the European Union nor the granting authority can be held responsible for them.
	D.B. acknowledges financial support from the Spanish Ministry of Science and Innovation (MICINN) through the Spanish State Research Agency, under Severo Ochoa Centres of Excellence Programme 2025-2029 (CEX2024001442-S).

	\appendix
	
	\section{Cutoff frequency}\label{app:cutoff_frequency}
	
	In this appendix, we compute the time scale $t_\star$ that appears as a regulator in the energy-momentum tensor in Eq.~\eqref{eq:Tmunu}. We consider a scenario in which a heavy particle of mass $M$ decays into two massless daughter particles. The daughter particles have momenta of order $M$, while the surrounding plasma off which they scatter has a temperature $T<M$.
	
	The time scale $t_\star$ corresponds to the mean free time (or equivalently, the mean free path) required for the initial decay products to lose an order-one fraction of their momentum, cf. Eq.~\eqref{eq:Tmunu}.
	This time scale can be estimated using results from the quark-gluon plasma literature. For example, the energy loss $\epsilon$, per unit length  $x$, of a muon propagating through a QED plasma is given by~\cite{Peigne:2007sd}
	\begin{eqnarray}
		\frac{d\epsilon}{dx}= \frac{e^4 T^2}{48 \pi} {\rm ln}\left(\frac{3M}{e^2 T}\right),
		\label{eq:dEdx}
	\end{eqnarray}
	where $e$ is a coupling constant. For our estimate, we consider only elastic scatterings with thermal electrons and positrons and neglect Compton scattering. We focus on a QED plasma because our goal is merely to estimate the mean free path. A more sophisticated estimate that includes QCD effects can be obtained by replacing $e^2\to 4\pi \alpha_s$ and adding a slightly different pre-factor, cf. Ref. \cite{Peigne:2008nd}. Note, however, that at high temperatures the SM gauge couplings are all on the same order $e\approx 0.5$, so our estimate is sufficiently accurate for our purposes.
	
	The inverse mean free path is now given by:
	\begin{eqnarray}
		\omega_\star = \frac{e^4 T^2}{48\pi\,M}\,{\rm ln}\left(\frac{3M}{e^2 T}\right)\,\frac{\zeta(3)}{\pi^2}\,g_{*{\rm SM},n},
		\label{eq:omega_star_QED}
	\end{eqnarray}
	where the last factor takes into account that there are multiple particle species in the thermal plasma, i.e. the number density is $n=\zeta(3)/\pi^2\, T^3 g_{*{\rm SM},n}$, where $g_{*{\rm SM},n}=95.5$.


	Redshifting $\omega_\star$ from the time where the particle decay happens to today, we obtain 
	\begin{eqnarray}
		\omega_{\star}^0\approx 4.0\times 10^7\,{\rm Hz}\,\left(\frac{T}{10^{12}\,{\rm GeV}}\right)\,\left(\frac{10^{14}\,{\rm GeV}}{M}\right)\,\left(\frac{e}{0.5}\right)^4,
		\label{eq:omega_TM_numbers}
	\end{eqnarray}
	where in order to get the correct pre-factor we have also evaluated the logarithm that appears in $\omega_\star^0$ at $T_D=10^{12}$ GeV, $M=10^{14}$ GeV and $e=0.5$, cf. Eq.~\eqref{eq:omega_star_QED}. We have also assumed that no entropy is produced after the decay.

	We compare $\omega_\star$ to the Hubble rate
	\begin{eqnarray}
		H = \frac{T^2}{m_{\rm pl}} \sqrt{\frac{8\pi^3\, g_{*\rm SM}}{90}},
		\label{eq:omega_TM_H}
	\end{eqnarray}
	where $H$ is the Hubble rate at the moment of the decay. In Eq.~\eqref{eq:omega_TM_H} we assume a radiation-dominated universe.  Plugging in numbers and redshifting to today, we obtain:
	\begin{eqnarray}
		\frac{H}{1+z}=1.6\times 10^5\,{\rm Hz}\,\left(\frac{T}{10^{12}\,{\rm GeV}}\right).
		\label{eq:omega_TM_H_numbers}
	\end{eqnarray}
	Comparing Eqs.~\eqref{eq:omega_star_QED} and~\eqref{eq:omega_TM_H}, we see that the Hubble rate is parametrically smaller by a factor of $M/m_{\rm pl}$. However, due to the small coupling coefficient $\sim e^4$, the scattering-induced cutoff frequency can become smaller for sufficiently large $M$, cf. Eqs.~\eqref{eq:omega_TM_numbers} and~\eqref{eq:omega_TM_H_numbers}. Alternatively, in weakly coupled BSM theories, where the coupling can be much smaller than in the Standard Model, or if one of the decay products is a hard graviton (as in the non-linear memory effect), the coupling can become so small that the scattering-induced cutoff falls below the Hubble scale.

	\section{Metric perturbation with dipole approximation}\label{app:metric_diploe_approximation}
	
	In order to derive the expression for the metric perturbation in the dipole approximation, cf. Eq.~\eqref{eq:humnubar}, we need to insert the energy-momentum tensor from Eq.~\eqref{eq:Tmunu_Fourier} into the expression for the metric perturbation given in Eq.~\eqref{eq:hmunu_far_field_final}. We encounter two integrals that we solve by closing the integration contour either in the upper or lower half-plane.

	The first integral,
	\begin{eqnarray}
		\int_{-\infty}^{\infty}d\omega \frac{1}{\omega-\bar{\omega}} e^{-i\omega (t-r)},
	\end{eqnarray}
	has a pole at $\bar{\omega}=i/\tau$. If $t-r>0$, we close the integration region in the lower half of the complex plane, and if $t-r<0$ we close it in the upper half. The integral is zero if $t-r>0$. It is only non-zero if $t-r<0$, and therefore the integral must be proportional to $\Theta(r-t)$. 
	
	We evaluate the integral in the upper half-plane with the residue theorem
	\begin{eqnarray}
		\int_{-\infty}^{\infty}d\omega \,  \frac{1}{\omega-\bar{\omega}} e^{-i\omega (t-r)}=2 \pi i e^{ (t-r)/\tau}\Theta(r-t).
	\end{eqnarray}

	The second integral that we need to evaluate is:
	\begin{eqnarray}
		\int_{-\infty}^{\infty}d\omega\, e^{-i\omega (t-r)}\frac{1}{\omega-\omega\hat{\bm{x}}\cdot \hat{\bm{n}}_a-\frac{-i}{t_\star}}=\frac{1}{1-\hat{\bm{x}}\cdot \hat{\bm{n}}_a}\int_{-\infty}^{\infty}d\omega\, e^{-i\omega (t-r)}\frac{1}{\omega-\bar{\omega}},
	\end{eqnarray}
	where we have defined $\bar{\omega}=-i/(t_\star(1-\hat{\bm{x}}\cdot \hat{\bm{n}}_a))$, which is now in the lower half-plane of the complex plane. Therefore, the integral is now proportional to $\Theta(t-r)$.
	Using the residue theorem gives us:
	\begin{eqnarray}
		\int_{-\infty}^{\infty} d\omega e^{-i\omega (t-r)}\frac{1}{\omega-\bar{\omega}} 
		= -2 \pi i  e^{- (t-r)/(t_\star(1-\hat{\bm{x}}\cdot \hat{\bm{n}}_a))}\Theta(t-r).
	\end{eqnarray}
	Putting everything together, we obtain the expression in Eq.~\eqref{eq:humnubar}.
	The calculation shows that introducing finite $\tau$ and $t_\star$ makes the integrals well defined. Without these regulators, one would have to introduce them ad hoc to obtain the correct causal behavior of the metric perturbation.

	\section{Gravitational energy spectrum from the energy momentum tensor}\label{app:Tmunu_gravity}
	
	In this appendix, we derive Eq.~\eqref{eq:dEdomega_main_text}, which gives the gravitational energy spectrum for a specified energy-momentum tensor, such as one describing the decay of a particle into other particles. Even if this is a textbook result, cf. \cite{Maggiore,Weinberg_Gravitation}, we discuss it here to highlight the different approximations of the final result. We begin with the gravitational energy-momentum tensor~\cite{Weinberg_Gravitation}:
	\begin{eqnarray}
		t_{\mu\kappa}(\bm{x},t)=\frac{1}{8\pi G}\left[R^{(2)}_{~~~\mu\kappa}-\frac{1}{2}\eta_{\mu\kappa}\eta^{\rho\sigma}R^{(2)}_{~~~\rho\sigma}\right],
	\end{eqnarray}
	where we used that the first order Ricci tensor $R^{(1)}_{~~\mu\kappa}$ vanishes. To see this, we expand the metric $g_{\mu\nu}=\eta_{\mu\nu}+h_{\mu\nu}$ in the Einstein equations $R_{\mu\nu}-\frac{1}{2}g_{\mu\nu}R=-8\pi G T^{\rm matter}_{\mu\nu}$, which at linear order yields $R^{(1)}_{\mu\nu}-\frac{1}{2}\eta_{\mu\nu}R^{(1)}=0$. Taking the trace gives $R^{(1)}=0$, which in turn implies $R^{(1)}_{\mu\nu}=0$.
	The second-order Ricci tensor is:
	\begin{eqnarray}
		R^{(2)}_{~~\mu\kappa}&=&-\frac{1}{2} h^{\lambda\nu}\left[\partial_\kappa\partial_\mu h_{\lambda\nu}-\partial_\kappa\partial_\lambda h_{\mu\nu}-\partial_\nu\partial_\mu h_{\lambda\kappa}+\partial_\nu\partial_\lambda h_{\mu\kappa}\right]
		+
		\frac{1}{4}\left[2\partial_\nu h^\nu_{~\sigma}-\partial_\sigma h\right]\, \left[\partial_\kappa h^\sigma_{~\mu}+\partial_\mu h^\sigma_{~\kappa}-\partial^\sigma h_{\mu\kappa}\right]\nonumber\\
		&~&
		-\frac{1}{4}\left[\partial_\lambda h_{\sigma\kappa}+\partial_\kappa h_{\sigma\lambda}-\partial_\sigma h_{\lambda\kappa}\right]\, \left[\partial^\lambda h^\sigma_{~\mu}+\partial_\mu h^{\sigma\lambda}-\partial^\sigma h^\lambda_{~\mu}\right].
	\end{eqnarray}
	We calculate the total gravitational energy $E$ flowing through the surface of a sphere centered at the origin.
	\begin{eqnarray}
		E= \int_{-\infty}^{\infty} dt \, \int_{S} dA^i \, t_{0i}(\bm{x},t),
		\label{eq:E}
	\end{eqnarray}
	where $S_r$ denotes a unit sphere and its surface element is given by $dA_i=\hat{\bm{e}}_{r,i} r^2 \,d\Omega$ with $d\Omega=\sin\theta \, d\theta \, d\phi$ and $\hat{\bm{e}}_r=(\sin\theta\,\cos\phi,\sin\theta\,\sin\phi,\cos\theta)$. Note that $dA^i=-dA_i$.

	To simplify the calculation, we evaluate the energy-momentum tensor in the so-called transverse-traceless (TT) gauge. Since the total emitted energy is gauge invariant, the final result should be independent of the gauge in which it is computed. We do not attempt to demonstrate this explicitly here; verifying the equivalence of the result in different gauges would be an interesting subject for future work. The TT gauge is obtained by imposing the De Donder gauge condition together with the additional requirement that the trace of the metric perturbation vanishes.
	\begin{eqnarray}
		\partial^\mu h^{{\rm TT}}_{ \mu\nu}=0,~~~ h^{\rm TT}=0,~~~h^{{\rm TT}}_{0\mu}=0.
		\label{eq:TT_conditions}
	\end{eqnarray}
	Far away from the source, it is always possible to choose a frame in which the metric perturbation satisfies the TT-gauge conditions. Since a symmetric $4\times 4$ matrix has ten independent parameters, the constraints in Eq.~\eqref{eq:TT_conditions} reduce the number of independent components to two. Note that the relation $\partial^\mu h^{{\rm TT}}_{ \mu0}=0$ is automatically fulfilled because we impose $h^{{\rm TT}}_{0\mu}=0$.
	We will explicitly indicate if tensors are evaluated in TT gauge or not. The transverse-traceless part of a tensor is extracted by the projection operator $\Lambda_{ijab}$:
	\begin{eqnarray}
		h^{\rm TT}_{ij}(\bm{k},\omega)=\Lambda_{ijab}(\hat{\bm{k}}) h_{ab}(\bm{k},\omega),
		\label{eq:hTT_def} 
	\end{eqnarray}
	with $\Lambda_{ijab}(\hat{\bm{k}})=P_{ia}P_{jb}-1/2\,P_{ij}P_{ab}$ and $P_{ij}=\delta_{ij}-\hat{k}_i\hat{k}_j$. Note that $\Lambda_{iiab}=0=\Lambda_{ijaa}$ (sum over $i$ and $a$). Furthermore $\Lambda_{ijab}\Lambda_{ablk}=\Lambda_{ijlk}$ and $k^i\Lambda_{ijab}(\hat{\bm{k}})=0=k^j\Lambda_{ijab}(\hat{\bm{k}})$. The definition of $h^{\rm TT}_{ij}(\bm{k},\omega)$ in Fourier space leads of course to an $h^{\rm TT}_{ij}(\bm{x},t)$ in real space that satisfies the constraints in Eq.~\eqref{eq:TT_conditions}.

	Before evaluating $R^{(2)}_{~~\mu\kappa}$ in the TT gauge, we note that we can always shuffle around partial derivatives at the cost of a sign change, for example $h^{\lambda\nu}\partial_\kappa\partial_\lambda h_{\mu\nu}=-\partial_\lambda h^{\lambda\nu}\partial_\kappa h_{\mu\nu}$. 
	To see this, one makes use of the far-field approximation of the metric perturbation, cf. Eq.~\eqref{eq:hmunu_far_field_final}. Furthermore, since we are interested in the total emitted energy, all expressions should be understood as being integrated over time, $\int_{-\infty}^{\infty}dt$. Upon substituting the Fourier representation of the metric perturbation, this integral yields a factor of $2\pi\delta(\omega-\omega')$. Another important observation is that, in the large $r$ limit, the derivative acting on the part of the metric perturbation proportional to $\sim T_{\mu\nu}(\bm{k}=\omega\hat{\bm{x}},\omega)/r$ can be neglected.

	Now evaluating $R^{(2)}_{\mu\nu}$ in the TT gauge and using the ``partial integration'' relations, we find:
	\begin{eqnarray}
		R^{(2)}_{~~\mu\kappa}&=&\frac{1}{4} \partial_\kappa h^{{\rm TT}\,\lambda\nu}\partial_\mu h_{\lambda\nu}^{\rm TT},
	\end{eqnarray}
	where we have also used that far from the source the metric perturbation satisfies $\partial_\mu\partial^\mu h^{\rm TT}_{\alpha\beta}=\partial_\mu\partial^\mu \bar{h}^{\rm TT}_{\alpha\beta}=0$. It then follows that $\eta^{\rho\sigma}R^{(2)}_{~~\rho\sigma}= 0$ and therefore
	\begin{eqnarray}
		t_{\mu\kappa}(\bm{x},t)=\frac{1}{32\pi G} \partial_\kappa h^{{\rm TT}\,\lambda\nu}\partial_\mu h_{\lambda\nu}^{\rm TT}.
	\end{eqnarray}
	The total emitted energy is:
	\begin{eqnarray}
		E =
		-\frac{1}{32\pi G}\,\int_{-\infty}^{\infty} dt' \, \int_{S} d\Omega \, \hat{\bm{e}}_{rl}\, r^2 \,  \dot{h}^{{\rm TT}\,ij}\partial_l h_{ij}^{\rm TT},
		\label{eq:E_tot}
	\end{eqnarray}
	where
	\begin{eqnarray}
		h_{ij}^{\rm TT}(\bm{x},t) = \frac{4G}{r}\,  \int\frac{d\omega}{2\pi} e^{-i\omega (t-r)} T^{\rm TT}_{ij}(\bm{k}=\omega\hat{\bm{x}},\omega) .
		\label{eq:hijTT_far_field}
	\end{eqnarray}
	The spatial derivative $\hat{\bm{e}}_{rl} \partial_l = \partial_r$ in Eq.~\eqref{eq:E_tot} acting on the metric perturbation is:
	\begin{eqnarray}
		\partial_r h_{ij}^{\rm TT}= h_{ij}^{\rm TT}/r
		+\frac{4G}{r}\,  \int\frac{d\omega}{2\pi} i\omega e^{-i\omega (t-r)} T^{\rm TT}_{ij}(\bm{k}=\omega\hat{\bm{x}},\omega).
	\end{eqnarray}
	We can neglect the first term because it scales as $ 1/r^2$ whereas the second term scales as $ 1/r$. Furthermore the second term can be rewritten as $\partial_r h_{ij}^{\rm TT}= -\partial_t h_{ij}^{\rm TT}$.
	Substituting this into Eq.~\eqref{eq:E_tot} yields:
	\begin{eqnarray}
		E 
		=
		\frac{1}{32\pi G}\,\int_{-\infty}^{\infty} dt' \, \int_{S} d\Omega \,  r^2 \,  \dot{h}^{{\rm TT}\,ij} \dot{h}_{ij}^{\rm TT}.
		\label{eq:E_tot_final}
	\end{eqnarray}
	Next, plugging in the expression for the metric perturbation from Eq.~\eqref{eq:hijTT_far_field} gives:
	\begin{eqnarray}
		E 
		=
		\frac{G}{2\pi }\, \, \int_{S} d\Omega \,   \,  
		\int_0^\infty\frac{d\omega}{2\pi} 2 \omega^2     
		T^{\rm TT}_{ij}(\bm{k}=\omega\hat{\bm{x}},\omega) 
		T^{{\rm TT}\, ij *}(\bm{k}=\omega\hat{\bm{x}},\omega) ,
	\end{eqnarray}
	where we have taken the complex conjugate of one $h_{ij}^{\rm TT}$, making use of the fact that the metric perturbation is real. We have also performed the time integral, which yields a factor $2\pi\delta(\omega-\omega')$ thereby allowing us to perform one of the frequency integrals. The extra factor of two was introduced because now we integrate the frequency integral from $0$ to $\infty$, while the frequency integral in Eq.~\eqref{eq:hijTT_far_field} runs from $-\infty$ to $\infty$.
	We can now read off the gravitational energy per frequency interval:
	\begin{eqnarray}
		\frac{dE}{d\omega}= 
		\frac{G}{2\pi^2 }\, \, \int_{S} d\Omega \,   \,  
		\omega^2     
		T^{\rm TT}_{ij}(\bm{k}=\omega\hat{\bm{x}},\omega) 
		T^{{\rm TT}\, ij *}(\bm{k}=\omega\hat{\bm{x}},\omega) .
		\label{eq:dEdomega}
	\end{eqnarray}
	Note that this is the one-sided energy spectrum, i.e., to obtain the total energy, one must integrate $dE/d\omega$ over $\omega$ from $0$ to $\infty$.
	A very similar expression was previously derived by Weinberg~\cite{Weinberg_Gravitation}; see the equation between Eqs. (10.4.16) and (10.4.17). The difference between Weinberg's expression and Eq.~\eqref{eq:dEdomega} is a factor $(2\pi)^2$. This difference arises from the Fourier transform convention: Weinberg uses $\int d\omega$, and we use $\int d\omega/(2\pi)$; see, for example,e Eq. (10.4.1) in Ref.~\cite{Weinberg_Gravitation}. Taking this into account, our result agrees with Weinberg's expression. There is, however, a slight difference in the projection tensor $\Lambda$. Our expression follows Ref.~\cite{Maggiore}. Writing all terms out and comparing it with Weinberg's definition shows that instead of the term $-2 k_j k_m \delta_{il}$ we obtain $- k_j k_m \delta_{il}- k_i k_l \delta_{jm}$. 
	The difference originates from the fact that Weinberg works in De Donder gauge and then uses energy-momentum conservation to express everything through the spatial components of the energy-momentum tensor. We expect that both choices for the projection tensor are equivalent.

	Many references emphasize averaging the energy-momentum tensor over a characteristic time or length scale. In our case, however, we are interested in the total emitted energy. The required integration over time effectively performs this averaging automatically. See also the discussion in Sec.~10.4 of Ref.~\cite{Weinberg_Gravitation}.

	\section{Derivation of cosmological stochastic gravitational energy spectrum from many events}\label{app:phinney}

	In this appendix, we compute the cosmological stochastic gravitational energy spectrum generated from many events in the early universe. One such event could, for example, be the decay of a particle. As we have emphasized in different places, if the decay products are absorbed at rates faster that the expansion rate, one can consider the injection of energy as instantaneous. We assume this hypothesis in this work, and leave any extension to future work.  Furthermore, the gravitational-wave energy spectrum produced by a single decay need not be monochromatic. We denote the number density of decaying particles by $n(z)$. Note that $n$ is defined with respect to the physical volume rather than the co-moving volume.  The final equation that we derive was previously obtained by Phinney~\cite{Phinney:2001di} in the context of stochastic gravitational wave backgrounds from a population of binary black hole mergers. Here, we re-derive this result while carefully stating all underlying assumptions. In the main text, we then apply the result to the case of particle decays in the early universe. Throughout this appendix, $t$ denotes cosmological time.

	Consider a co-moving volume element $dV$. During the time interval $(t,t+dt)$, the number of gravitons produced within $dV$ with frequencies in the interval  $(\omega,\omega+d\omega)$ is given by:
	\begin{eqnarray}
		\frac{dN_h}{dV\,dt\, d\omega} = \frac{dN_{\rm g}}{d\omega}\frac{dN}{dV\,dt},
		\label{eq:dNh_DIV_dVdtdomega}
	\end{eqnarray}
	where $dN_{\rm g}/d\omega$ is the number of gravitons from one single decay with frequencies in the interval $(\omega,\omega+d\omega)$ and $dN/(dV\,dt)$ denotes the number of particles in $dV$ that decay during the time interval $(t,t+dt)$. Note that we implicitly assumed in the above equations that the gravitational radiation emission from a single event is instantaneous. If this assumption were relaxed, one would need to account for the fact that decays occurring at earlier times could also contribute to the graviton production during the interval $(t,t+dt)$.

	An observer located at a co-moving distance $d_c$ from the volume element $dV$ will observe $dN_h^{0}$ gravitons that were emitted from $dV$ during the time interval $(t,t+dt)$ and have frequencies in the interval $(\omega,\omega+d\omega)$:
	\begin{eqnarray}
		\frac{dN_h^{0}}{dV\,dt\, d\omega} = \frac{dN_h}{dV\,dt\, d\omega} \frac{dA}{4\pi d_c^2},
		\label{eq:dNhobs_DIV_dtdomegadV}
	\end{eqnarray}
	where $dA$ represents the detector area of the observer. It is also defined in co-moving units; however, on the scale of an experiment, co-moving and proper distances are the same.

	The frequency interval $d\omega$ is measured in the rest frame of the volume element $dV$. An observer today would instead measure a frequency interval $(\omega_0,\omega_0+d\omega_0)$, and $d\omega_0$ is related to $d\omega$ through $d\omega= (1+z)d\omega_0$, where redshift $z$ serves as a proxy for the emission time. A larger redshift corresponds to an earlier emission time and consequently to a signal that has traveled a greater distance before reaching us today.
	A similar argument holds for the time interval. An observer at $dV$ measures the time interval $(t,t+dt)$, while for an observer today $dt$ translates to $dt_0=(1+z)dt$.
	
	We now calculate the gravitational energy density $d\rho_h^{0}$ contributed by radiation originating from the volume element $dV$ at co-moving distance $d_c$. We consider radiation emitted during the time interval $(t,t+dt)$ in the frequency range $(\omega,\omega+d\omega)$
	\begin{eqnarray}
		\frac{d\rho_h^{0}}{d\omega_0dV}=\frac{dN_h^{0}}{dt_0d\omega_0dV}\frac{1}{dA} \frac{\omega_0}{c},
		\label{eq:dNhobs_DIV_dttdtomegadV}
	\end{eqnarray}
	where $cdt_0\,dA$ is a volume element at the observer and $dN_h^{0}\,\omega_0$ is the gravitational energy in the frequency interval $(\omega_0,\omega_0+d\omega_0)$. We have written the factor of $c$ explicitly to emphasize that $cdt_0\,dA$ is a volume element. From this point onward we again set $c=1$. Note that the time interval $(t,t+dt)$ is not arbitrary. It is related to the co-moving distance $d_c$ of the volume element $dV$, since the radiation emitted from $dV$ must arrive at the observer precisely at the time of observation.

	Using $d\omega_0dt_0=d\omega dt$ and Eq.~\eqref{eq:dNhobs_DIV_dtdomegadV} we can re-write the RHS of Eq.~\eqref{eq:dNhobs_DIV_dttdtomegadV}:
	\begin{eqnarray}
		\frac{d\rho_h^{0}}{d\omega_0dV}=\frac{dN_h^{0}}{dtd\omega dV}\frac{1}{dA} \omega_0 =\frac{dN_h}{dV\,dt\, d\omega} \frac{1}{4\pi d_c^2} \omega_0.
	\end{eqnarray}
	This is the central expression for the energy density $d\rho_h^{0}$ that is coming from the volume element $dV$, which is located at a co-moving distance $d_c$ from an observer. The radiation was emitted during the time interval $(t,t+dt)$ and within the frequency range $(\omega,\omega+d\omega)$. Integrating over the entire co-moving volume yields:
	\begin{eqnarray}
		\frac{d\rho_h^{0}}{d\omega_0}=\int dd_c\,\frac{dN_h}{dV\,dt\, d\omega}  \omega_0,
	\end{eqnarray}
	where we have used that $dV = 4\pi d_c^2 dd_c$.
	We can use Eq.~\eqref{eq:dNh_DIV_dVdtdomega} for the first term under the integral
	\begin{eqnarray}
		\frac{d\rho_h^{0}}{d\omega_0}=\int dd_c\,\frac{dN_{\rm g}}{d\omega}\frac{dN}{dV\,dt}\,  \omega_0.
	\end{eqnarray}
	Next we insert $dN_{\rm g}/d\omega= dE/d\omega \times 1/\omega$, where $dE$ is the gravitational energy in the frequency range $(\omega,\omega+d\omega)$. Furthermore, we convert the integral to an integral over redshift:
	\begin{eqnarray}
		\frac{d\rho_h^{0}}{d\omega_0}=\int dz\,\frac{dd_c}{dz}\,\frac{dE}{d\omega}\frac{1}{\omega}\frac{dN}{dV\,dz}\, \frac{dz}{dt}\, \omega_0.
	\end{eqnarray}
	We use that $dz/dt = -H (1+z)$ and $dd_c/dz = -1/H$. Putting everything together, we obtain an expression for the gravitational energy spectrum:
	\begin{eqnarray}
		\Omega_{\rm h} =\frac{1}{\rho_c^0} \frac{d\rho_h^{0}}{d{\rm ln}\omega_0} = \frac{1}{\rho_c^0}  \int dz\,\,\frac{\omega}{(1+z)}\frac{dE}{d\omega}\frac{dN}{dV\,dz},
		\label{eq:phinney}
	\end{eqnarray}
	where $\rho_c^0$ is the critical energy density today. Eq.~\eqref{eq:phinney} is exactly the expression that was previously derived in Ref.~\cite{Phinney:2001di}.

	\section{Details for metric perturbation calculation}\label{app:details_metric}
	
	In this appendix, we derive the metric perturbation for the process in which a massive non-relativistic particle decays into two massless particles.  The solution to the wave equation, cf. Eq.~\eqref{eq:Wave_eq_hmunu}, is:
	\begin{eqnarray}
		\bar{h}_{\mu\nu}(\bm{x},t)=4 G\int d^3x'\frac{1}{|\bm{x}-\bm{x}'|}T_{\mu\nu}(\bm{x}',t-|\bm{x}-\bm{x}'|)=\bar{h}_{\mu\nu}^A+\bar{h}_{\mu\nu}^B+\bar{h}_{\mu\nu}^C,
	\end{eqnarray}
	where the index $A$ represents the contribution from the decaying particle, while $B,C$ label the contributions from the two decay products:
	\begin{eqnarray}
		\bar{h}_{\mu\nu}^A &=& 4 G\int d^3x'\frac{1}{|\bm{x}-\bm{x}'|}
		\frac{p_\mu p_\nu}{p^0}\delta(\bm{x}')\Theta(-t+|\bm{x}-\bm{x}'|),\\
		\bar{h}_{\mu\nu}^B&=&4 G\int d^3x'\frac{1}{|\bm{x}-\bm{x}'|}\frac{p_{1\mu} p_{1\nu}}{p_1^0}\delta(\bm{x}'-\bm{x}_{1}(t-|\bm{x}-\bm{x}'|))\Theta(t-|\bm{x}-\bm{x}'|),\\
		\bar{h}_{\mu\nu}^C&=&4 G\int d^3x'\frac{1}{|\bm{x}-\bm{x}'|}\frac{p_{2\mu} p_{2\nu}}{p_2^0}\delta(\bm{x}'-\bm{x}_{2}(t-|\bm{x}-\bm{x}'|))\Theta(t-|\bm{x}-\bm{x}'|),
	\end{eqnarray}
	where $\bm{x}_{1,2}=\hat{\bm{n}}_{1,2}t$ and we chose $\hat{\bm{n}}_{1,2}=\pm \hat{\bm{e}}_z$ without loss of generality.
	The expression for $\bar{h}^A_{\mu\nu}$ is:
	\begin{eqnarray}
		\bar{h}_{\mu\nu}^A = 4 G\frac{1}{r}\frac{p_\mu p_\nu}{p^0}\Theta(-t+|\bm{x}|).
		\label{eq:hmunubarA}
	\end{eqnarray}
	In order to evaluate the integral in $\bar{h}^B_{\mu\nu}$ we need to look at the delta function:
	\begin{eqnarray}
		\delta\left(\bm{x}'-\bm{x}_{1}(t-|\bm{x}-\bm{x}'|)\right)=
		\delta(x')\delta(y')\delta(z'-t+|\bm{x}-\bm{x}'|).
	\end{eqnarray}  
	This can be simplified:
	\begin{eqnarray}
		\delta\left(\bm{x}'-\bm{x}_{1}(t-|\bm{x}-\bm{x}'|)\right)=\delta(x')\delta(y')\frac{\delta(z'-z_0')}{|f'(z_0')|},
	\end{eqnarray}
	where
	\begin{eqnarray}
		f(z')=z'-t+|\bm{x}-\bm{x}'|=z'-t+\sqrt{x^2+y^2+(z-z')^2},
	\end{eqnarray}
	$f'$ denotes the derivative of $f$ and $z_0'$ is the zero
	\begin{eqnarray}
		z_0'=\frac{t^2-\bm{x}^2}{2(t-z)}.
		\label{eq:zp0_B}
	\end{eqnarray}

	Similarly we can simplify the delta function that appears in $\bar{h}^C_{\mu\nu}$:
	\begin{eqnarray}
		\delta(\bm{x}'-\bm{x}_{2}(t-|\bm{x}-\bm{x}'|))=\delta(x')\delta(y') \frac{\delta(z'-z_0')}{|f'(z_0')|},
	\end{eqnarray}
	where
	\begin{eqnarray}
		f(z')=z'+t-|\bm{x}-\bm{x}'| =z'+t-\sqrt{x^2+y^2+(z-z')^2}
	\end{eqnarray}
	and
	\begin{eqnarray}
		z_0'=\frac{t^2-\bm{x}^2}{-2(t+z)}.
		\label{eq:zp0_C}
	\end{eqnarray}

	Evaluating the integrals gives us for $\bar{h}_{\mu\nu}^B$
	\begin{eqnarray}
		\bar{h}_{\mu\nu}^B=4 G\frac{p_{1\mu} p_{1\nu}}{p_1^0}\frac{1}{\sqrt{x^2+y^2+(z-z_0')^2}+z_0'-z}\Theta(t-|\bm{x}-\bm{x}'|)
	\end{eqnarray}
	with $\bm{x}'=(0,0,z_0')$ and $z_0'$ from Eq.~\eqref{eq:zp0_B}.
	For $\bar{h}_{\mu\nu}^C$ we obtain:
	\begin{eqnarray}
		\bar{h}_{\mu\nu}^C=4 G\frac{p_{2\mu} p_{2\nu}}{p_2^0}\frac{1}{\sqrt{x^2+y^2+(z-z_0')^2}+z-z_0'}\Theta(t-|\bm{x}-\bm{x}'|)
	\end{eqnarray}
	with $\bm{x}'=(0,0,z_0')$ and $z_0'$ from Eq.~\eqref{eq:zp0_C}.
	
	We can simplify the denominators of $\bar{h}_{\mu\nu}^B$ and $\bar{h}_{\mu\nu}^C$
	\begin{eqnarray}
		\text{denominator in B} = \begin{cases}
			t-z & \text{for } t-z>0\\
			-(x^2+z^2)/(t-z) & \text{for } t-z<0
		\end{cases}
	\end{eqnarray}
	\begin{eqnarray}
		\text{denominator in C} = \begin{cases}
			t+z & \text{for } t+z>0\\
			-(x^2+z^2)/(t+z) & \text{for } t+z<0
		\end{cases}
	\end{eqnarray}
	The arguments in the theta-functions can also be simplified.
	\begin{eqnarray}
		\text{$\Theta$-function argument B} = t-\left|\bm{x}-\left(0,0,\frac{t^2-\bm{x}^2}{2(t-z)}\right)\right| = \begin{cases}
			\frac{t^2-\bm{x}^2}{2(t-z)} & \text{for } t-z>0\\
			\frac{-x^2-y^2-(z-2t)^2+t^2}{2(z-t)} & \text{for } t-z<0
		\end{cases}
	\end{eqnarray}
	Let us first look at the case $t-z>0$ in the above equation. In this scenario, only the numerator is relevant because we are looking at an argument of a theta-function. We can write $\Theta(t^2-r^2)=\Theta(t-r)$, and in the following we refer to the regime $t>r$ as the \textit{causal sphere}, since it contains all information emitted from particle $B$ that propagates at or below the speed of light.  Now consider the case $t-z<0$. As before, the denominator is always positive, so the sign of the theta function argument is solely determined by the numerator. The numerator describes a sphere of radius $t$ centered at $z=2t$. Inside this sphere, the argument of the theta function is positive. We refer to this as the \textit{acausal sphere} since information originating from particle $B$ cannot enter it without propagating faster than light.
	The acausal sphere is an artifact of the ansatz used for the trajectories of the massless particles. To avoid this issue, the particle trajectories should be modified so that they vanish for negative times, i.e., $\bm{x}_1=(0,0,t)\Theta(t)$ and $\bm{x}_2=(0,0,-t)\Theta(t)$. That way we would get rid of the acausal solution. 
	In the following, we therefore drop the acausal solutions because they are clearly unphysical. Likewise, we discard the corresponding unphysical contributions to $\bar{h}^C_{\mu\nu}$.
	The physical solutions for $\bar{h}^B_{\mu\nu}$ and $\bar{h}^C_{\mu\nu}$ are:
	\begin{eqnarray}
		\bar{h}_{\mu\nu}^B =4 G\frac{p_{1\mu} p_{1\nu}}{p_1^0}\frac{1}{t-z}\Theta(t-|\bm{x}|),\\
		\bar{h}_{\mu\nu}^C =4 G\frac{p_{2\mu} p_{2\nu}}{p_2^0}\frac{1}{t+z}\Theta(t-|\bm{x}|).
	\end{eqnarray}

	The interpretation of the metric perturbations is as follows: an observer at $\bm{x}_O$ initially experiences only the contribution $h^A_{\mu\nu}$. Once $t=|\bm{x}_O|$, the decay becomes causally observable, and the observer begins to feel the effects of $h^B_{\mu\nu}$ and $h^C_{\mu\nu}$. While the $A$ term is time independent (up to the time dependence in the theta-function), the $B$ and $C$ terms decay for $t\to\infty$.
	
	In deriving the above expressions, we have assumed that the outgoing particles (1 and 2) propagate along the positive and negative z-directions, respectively. However, by rotational symmetry, the result generalizes to arbitrary directions $\hat{\bm{n}}=\hat{\bm{n}}_1=-\hat{\bm{n}}_2$:
	\begin{eqnarray}
		\bar{h}_{\mu\nu}^B &=& 4G\frac{1}{t-r\hat{\bm{x}}\cdot \hat{\bm{n}}_1} \frac{p_{1\mu} p_{1\nu}}{p^0_1}\,\Theta(t-r),\label{eq:hmunubarB}\\
		\bar{h}_{\mu\nu}^C &=& 4G\frac{1}{t-r\hat{\bm{x}}\cdot \hat{\bm{n}}_2} \frac{p_{2\mu} p_{2\nu}}{p^0_2}\,\Theta(t-r).\label{eq:hmunubarC}
	\end{eqnarray}

	\newpage
	\bibliographystyle{utcaps}
	\bibliography{references}
	
\end{document}